\pdfoutput=1
\RequirePackage{ifpdf}
\ifpdf 
\documentclass[pdftex]{sigma}
\else
\documentclass{sigma}
\fi

\numberwithin{equation}{section}
\usepackage[all]{xy}
\usepackage{multirow}

\numberwithin{theorem}{section}
\numberwithin{definition}{section}
\numberwithin{example}{section}

\DeclareMathOperator{\ord}{ord}

\begin{document}

\newcommand{\arXivNumber}{1502.00128}

\allowdisplaybreaks

\renewcommand{\thefootnote}{$\star$}

\renewcommand{\PaperNumber}{043}

\FirstPageHeading

\ShortArticleName{Structure Relations and Darboux Contractions for 2D 2nd Order Superintegrable Systems}

\ArticleName{Structure Relations and Darboux Contractions
\\
for 2D 2nd Order Superintegrable Systems\footnote{This paper is a~contribution to the Special Issue on Exact
Solvability and Symmetry Avatars in honour of Luc Vinet.
The full collection is available at
\href{http://www.emis.de/journals/SIGMA/ESSA2014.html}{http://www.emis.de/journals/SIGMA/ESSA2014.html}}}

\Author{Robin HEINONEN~$^\dag$, Ernest G.~KALNINS~$^\ddag$, Willard MILLER Jr.~$^\dag$ and Eyal SUBAG~$^\S$}

\AuthorNameForHeading{R.~Heinonen, E.G.~Kalnins, W.~Miller Jr. and E.~Subag}

\Address{$^\dag$~School of Mathematics, University of Minnesota, Minneapolis, Minnesota 55455, USA}
\EmailD{\href{mailto:heino027@umn.edu}{heino027@umn.edu}, \href{mailto:miller@ima.umn.edu}{miller@ima.umn.edu}}
\URLaddressD{http://www.ima.umn.edu/~miller/}

\Address{$^\ddag$~Department of Mathematics, University of Waikato, Hamilton, New Zealand}
\EmailD{\href{mailto:math0236@waikato.ac.nz}{math0236@waikato.ac.nz}}

\Address{$^\S$~School of Mathematical Science, Tel Aviv University, Tel Aviv 69978, Israel}
\EmailD{\href{mailto:eyalsubag@gmail.com}{eyalsubag@gmail.com}}

\ArticleDates{Received February 03, 2015, in f\/inal form May 30, 2015; Published online June 08, 2015}

\Abstract{Two-dimensional quadratic algebras are generalizations of Lie algebras that include the symmetry algebras of
2nd order superintegrable systems in 2 dimensions as special cases.
The superintegrable systems are exactly solvable physical systems in classical and quantum mechanics.
Distinct superintegrable systems and their quadratic algebras can be related by geometric contractions, induced by
In\"onu--Wigner type Lie algebra contractions.
These geometric contractions have important physical and geometric meanings, such as obtaining classical phenomena as
limits of quantum phenomena as~${\hbar}\to 0$ and non\-re\-la\-tivistic phenomena from special relativistic as $c\to \infty$,
and the derivation of the Askey scheme for obtaining all hypergeometric orthogonal polynomials as limits of Racah/Wilson
polynomials.
In this paper we show how to simplify the structure relations for abstract nondegenerate and degenerate quadratic
algebras and their contractions.
In earlier papers we have classif\/ied contractions of 2nd order superintegrable systems on constant curvature spaces and
have shown that all results are derivable from free quadratic algebras contained in the enveloping algebras of the Lie
algebras $e(2,{\mathbb C})$ in f\/lat space and $o(3,{\mathbb C})$ on nonzero constant curvature spaces.
The quadratic algebra contractions are induced by generalizations of In\"on\"u--Wigner contractions of these Lie
algebras.
As a~special case we obtained the Askey scheme for hypergeometric orthogonal polynomials.
After constant curvature spaces, the 4 Darboux spaces are the 2D manifolds admitting the most 2nd order Killing tensors.
Here we complete this theoretical development for 2D superintegrable systems by showing that the Darboux superintegrable
systems are also characterized by free quadratic algebras contained in the symmetry algebras of these spaces and that
their contractions are also induced by In\"on\"u--Wigner contractions.
We present tables of the contraction results.}


\Classification{22E70; 16G99; 37J35; 37K10; 33C45; 17B60}

\renewcommand{\thefootnote}{\arabic{footnote}} \setcounter{footnote}{0}

\section{Introduction}

We def\/ine an abstract {\it nondegenerate (quantum) quadratic algebra} as a~noncommutative associative algebra generated
by linearly independent operators~$H$, $L_1$, $L_2$ such that~$H$ is in the center, $R=[L_1,L_2]\ne 0$ and the following
relations hold:
\begin{gather*}
[L_j,R]=\sum\limits_{0\leq e_1+e_2+e_3\leq 2} M^{(j)}_{e_1,e_2,e_3} \big\{L_1^{e_1}, L_2^{e_2}\big\} H^{e_3},
\qquad
e_k\geq 0,
\qquad
L_k^0=I,
\end{gather*}
for some $M^{(j)}_{e_1, e_2, e_3}\in \mathbb{C}$, and where $[A,B]=AB-BA$ is the commutator and
$\{L_1,L_2\}=L_1L_2+L_2L_1$ is the symmetrizer.
Also the operator $R^2$ is contained in the algebra of symmetrized products:
\begin{gather*}
R^2 -F\equiv R^2-\sum\limits_{0\leq e_1+e_2+e_3\leq 3} N_{e_1,e_2,e_3} \big\{L_1^{e_1}, L_2^{e_2}\big\} H^{e_3} = 0
\end{gather*}
for some $N_{e_1, e_2, e_3}\in \mathbb{C}$.

An abstract {\it degenerate $($quantum$)$ quadratic algebra} is a~noncommutative associative algebra generated by linearly
independent operators $X$, $H$, $L_1$, $L_2$ such that~$H$ is in the center and the following relations hold:
\begin{gather*}
[X,L_j]=\sum\limits_{0\leq e_1+e_2+e_3+e_4\leq 1} P^{(j)}_{e_1,e_2,e_3, e_4} L_1^{e_1}L_2^{e_2}H^{e_3}X^{e_4},
\qquad
j=1,2,
\end{gather*}
for some $P^{(j)}_{e_1,e_2,e_3, e_4} \in \mathbb{C}$.
The commutator $[L_1,L_2]$ is expressed as
\begin{gather*}
[L_1,L_2]=\sum\limits_{0\leq e_1+e_2+e_3+e_4\leq 1} Q_{e_1,e_2, e_3, e_4}\big\{L_1^{e_1}L_2^{e_2},X\big\}H^{e_3}X^{2e_4}
\end{gather*}
for some $Q_{e_1,e_2, e_3, e_4} \in \mathbb{C}$.
Finally, there is the relation:
\begin{gather*}
G\equiv\sum\limits_{0\leq e_1+e_2+e_3+e_4\leq 2} S_{e_1, e_2, e_3, e_4}\big\{L_1^{e_1},L_2^{e_2}, X^{2e_4}\big\}H^{e_3}=0, \qquad  X^0=H^0=I,
\end{gather*}
for some $S_{e_1,e_2, e_3, e_4} \in \mathbb{C}$ and where $\{L_1^{e_1},L_2^{e_2}, X^{2e_4}\}$ is the 6-term symmetrizer
of three operators.

For both quantum quadratic algebras there is a~natural grading such that the opera\-tors~$H$,~$L_j$ are 2nd order,~$X$ is 1st
order and
\begin{gather}
\label{orderrelations}
\ord([A,B])\le \ord(A)+\ord(B)-1,
\qquad
\ord(AB)=\ord(A)+\ord(B),
\\
\nonumber
\ord(I)=0,
\qquad
\ord(A+B)=\max\{\ord(A), \ord(B)\},
\qquad
\ord(cA)=\ord(A),
\end{gather}
for operators $A$, $B$, identity operator~$I$ and scalar~$c$, with $A\ne -B$, $A, B \ne 0$, and $c\ne 0$.
Thus~$R$ is usually 3rd order, expression~$G$ is 4th order and~$F$ is 6th order.
The f\/ield of scalars can be either ${\mathbb R}$ or ${\mathbb C}$.

There is an analogous quadratic algebra structure for Poisson algebras.
An abstract {\it non\-de\-ge\-nerate $($classical$)$ quadratic algebra} is a~Poisson algebra with functionally independent
genera\-tors~${\cal H}$, ${\cal L}_1$, ${\cal L}_2$ such that all generators are in involution with ${\cal H}$ and the following
relations hold:
\begin{gather*}
\{{\cal L}_j,{\cal R}\}=\sum\limits_{0\leq e_1+e_2+e_3\leq 2} M^{(j)}_{e_1,e_2,e_3} {\cal L}_1^{e_1}{\cal
L}_2^{e_2}{\cal H}^{e_3},
\qquad
e_k\geq 0,
\qquad
{\cal L}_k^0=1,
\\
{\cal R}^2-{\cal F}\equiv {\cal R}^2 -\sum\limits_{0\leq e_1+e_2+e_3\leq 3} N_{e_1,e_2,e_3} {\cal L}_1^{e_1}{\cal
L}_2^{e_2}{\cal H}^{e_3} = 0
\end{gather*}
for some $M^{(j)}_{e_1, e_2, e_3}, N_{e_1, e_2, e_3}\in \mathbb{C}$.
An abstract {\it degenerate $($classical$)$ quadratic algebra} is a~Poisson algebra with linearly independent generators
${\cal X}$, ${\cal H}$, ${\cal L}_1$, ${\cal L}_2$ such that all generators are in involution with {\cal H} and obey structure
equations
\begin{gather}
\{{\cal X},{\cal L}_j\}=\sum\limits_{0\leq e_1+e_2+e_3+e_4\leq 1} P^{(j)}_{e_1,e_2,e_3, e_4} {\cal L}_1^{e_1}{\cal
L}_2^{e_2}{\cal H}^{e_3}{\cal X}^{e_4},
\qquad
j=1,2,
\nonumber
\\
\{{\cal L}_1,{\cal L}_2\}=\sum\limits_{0\leq e_1+e_2+e_3+e_4\leq 1} Q_{e_1,e_2, e_3, e_4} {\cal L}_1^{e_1}{\cal
L}_2^{e_2}{\cal X}{\cal H}^{e_3}{\cal X}^{2e_4},
\nonumber
\\
\label{Bracket2p}
{\cal G}\equiv\sum\limits_{0\leq e_1+e_2+e_3+e_4\leq 2} S_{e_1, e_2, e_3, e_4}{\cal L}_1^{e_1}{\cal L}_2^{e_2}{\cal X}^{2e_4}{\cal H}^{e_3}=0,
\qquad
{\cal X}^0={\cal H}^0=1
\end{gather}
for some $P^{(j)}_{e_1, e_2, e_3, e_4}$, $Q_{e_1, e_2, e_3, e_4}$, $S_{e_1, e_2, e_3, e_4}\in \mathbb{C}$.
There is a~grading for these quadratic algebras with properties analogous to~\eqref{orderrelations}, but with the
Poisson bracket instead of the commutator.

These quadratic algebra structures arise naturally in the study of classical and quantum superintegrable systems in two
dimensions and are key to the exact solvability of these systems,
e.g.,~\cite{BDK,DASK2007,EVAN,FORDY,Zhedanov1992a,Zhedanov1992b,VILE,MPW2013,TTW2001,SCQS}.
A~quantum 2D superintegrable system is an integrable Hamiltonian system on an $2$-dimensional
Riemannian/pseudo-Riemannian manifold with potential that admits $3$ algebraically independent partial dif\/ferential
operators commuting with~$H$, the maximum possible.
\begin{gather*}
H=\Delta+V,
\qquad
[H,L_j]=0,
\qquad
L_{3}=H, \qquad  j=1,2,3.
\end{gather*}
(In 2 dimensions we can always f\/ind Cartesian-like coordinates $x_1$, $x_2$ such that
\[
H=\frac{1}{\lambda(x_1,x_2)}\big(\partial_{x_1}^2+\partial_{x_2}^2\big)+V(x_1,x_2)
\] and we adopt these coordinates in the
following.) A~system is of order~$k$ if the maximum order of the symmetry operators $L_j$ (other than~$H$) is~$k$; all
such systems are known for $k=1,2$~\cite{KKM20041,KKM20041+1,KKM20041+2,KKM20041+3, Koenigs}.
Superintegrability captures the properties of quantum Hamiltonian systems that allow the Schr\"odinger eigenvalue
problem $H\Psi=E\Psi$ to be solved exactly, analytically and algebraically.
A~classical 2D superintegrable system is an integrable Hamiltonian system on an $2$-dimensional
Riemannian/pseudo-Riemannian manifold with potential that admit 3 functionally independent
phase space functions ${\cal H}$, ${\cal L}_1$, ${\cal L}_2$ in involution with~${\cal H}$, the maximum possible:
\begin{gather*}
{\cal H}=\frac{p_1^2+p_2^2}{\lambda({\bf x})}+V({\bf x}),
\qquad
\{{\cal H},{\cal L}_j\}=0,
\qquad
{\cal L}_{3}={\cal H}, \qquad  j=1,2,3,
\end{gather*}
expressed in local Cartesian-like coordinates $x_1$, $x_2$, $p_1$, $p_2$.
A~system is of order~$k$ if the maximum order of the constants of the motion $L_j$, $j\ne 3$, as polynomials in
$p_1$, $p_2$ is~$k$.
Again all such systems are known for $k=1,2$, and there is a~1-1 relationship
between classical and quantum 2nd order 2D
superintegrable systems~\cite{KKM20041,KKM20041+1,KKM20041+2,KKM20041+3}.

The possible superintegrable systems divide into four classes:
\begin{enumerate}\itemsep=0pt
\item \emph{First order systems}.
These are the (zero-potential) Laplace--Beltrami eigenvalue equations on constant curvature spaces.
The symmetry algebras close under commutation to form the Lie algebras $e(2,{\mathbb R})$, $e(1,1)$, $o(3,{\mathbb R})$
or $o(2,1)$.
Such systems have been studied in detail, using group theory methods, e.g.,~\cite{Miller, Talman}.
\item \emph{Free triplets}.
These are superintegrable systems with zero potential and all generators of 2nd order.
The possible spaces for which these systems can occur were classif\/ied by Koenigs~\cite{Koenigs}.They are: constant
curvature spaces (6 linearly independent 2nd order symmetries, 3 1st order), the four Darboux spaces (4 linearly
independent 2nd order symmetries, 1 1st order), and eleven 4-parameter Koenigs spaces (3 linearly independent 2nd order
symmetries, 0 1st order).
In most cases the symmetry operators will not generate a~quadratic algebra, i.e., the algebra will not close.
If the system generates a~nondegenerate quadratic algebra we call it a~{\it free quadratic triplet}.
\item \emph{Nondegenerate systems}~\cite{KKM20041,KKM20041+1,KKM20041+2,KKM20041+3}.
Here all symmetries are of 2nd order and the space of potentials is 4-dimensional:
\begin{gather*}
V({\bf x})= a_1V_{(1)}({\bf x})+a_2V_{(2)}({\bf x})+a_3V_{(3)}({\bf x})+a_4.
\end{gather*}
The symmetry operators generate a~nondegenerate
quadratic algebra with parameters $a_j$.
\item \emph{Degenerate systems}~\cite{KKMP2009}. There are 4 generators: one 1st order~$X$ and 3 second order $H$, $L_1$, $L_2$.
Here, $X^2$ is not contained in the span of $H$, $L_1$, $L_2$.
The space of potentials is 2-dimensional: $V({\bf x})= a_1V_{(1)}({\bf x})+a_2$.
The symmetry operators generate a~degenerate quadratic algebra with parameters $a_j$.
Relation~\eqref{Bracket2p} is an expression of the fact that 4 symmetry operators cannot be algebraically independent.
\end{enumerate}

Every degenerate superintegrable system occurs as a~restriction of the 3-parameter potentials (i.e., 4-dimensional
potential space) to 1-parameter ones, such that one of the symmetries becomes a~perfect square: $L=X^2$.
Here~$X$ is a~f\/irst order symmetry and a~new 2nd order symmetry appears so that this restriction admits more symmetries
than the original system.

Strictly speaking, a~nondegenerate 2D superintegrable system, both classical and quantum, is not a~single system but in
fact a~family of superintegrable systems parameterized by three parameters, $a_1$, $a_2$, $a_3$.
Similarly a~degenerate 2D superintegrable system, both classical and quantum, is a~family of superintegrable systems
parameterized by one parameter, $a_1$.

For a~quadratic algebra that comes from a~nondegenerate 2D superintegrable system (classical and quantum) the constants
$M^{(j)}_{e_1, e_2, e_3}$ and $N_{e_1, e_2, e_3}$ are polynomials in the parame\-ters~$a_1$,~$a_2$,~$a_3$ of degree
$2-e_1-e_2-e_3$ and $3-e_1-e_2-e_3$, respectively.
If all parameters $a_j=0$ the algebra is {\it free}.
For a~quadratic algebra that comes from a~degenerate 2D superintegrable system (classical and quantum) the constants
$P^{(j)}_{e_1, e_2, e_3, e_4}$, $Q_{e_1, e_2, e_3, e_4}$ and $S_{e_1, e_2, e_3, e_4}$ are polynomials in~$a_1$ of
degrees $1-e_1-e_2-e_3-e_4$, $1-e_1-e_2-e_3-e_4$ and $2-e_1-e_2-e_3-e_4$, respectively.
If all parameters $a_j=0$ these algebras are {\it free}.

Basic results that relate these superintegrable systems are the closure theorems:
\begin{theorem}
\label{theorem2}
A~free triplet, classical or quantum, extends to a~superintegrable system with potential if and only if it generates
a~free quadratic algebra $\tilde Q$.
\end{theorem}
\begin{theorem}
\label{theorem3}
A~superintegrable system, degenerate or nondegenerate, classical or quantum, with quadratic algebra~$Q$, is uniquely
determined by its free quadratic algebra $\tilde Q$.
\end{theorem}
These theorems were proved in~\cite{KM2014}, except for systems on the Darboux spaces which will be proved in this
paper.
The proofs are constructive: Given a~free quadratic algebra $\tilde Q$ one can compute the potential~$V$ and the
symmetries of the quadratic algebra~$Q$.
Thus as far as superintegrable systems on specif\/ic spaces are concerned, all information about the systems is contained
in the free quadratic algebras.

We will refer to quadratic algebras associated with superintegrable systems as {\it geometric}, a~subset of {\it
abstract} quadratic algebras.
For quadratic algebras associated with quantum superintegrable systems the order of~$A$ is its order as a~partial
dif\/ferential operator.
For quadratic algebras associated with classical superintegrable systems the order of $\cal A$ is its order as
a~polynomial in the momenta.
Although there is a~1-1 relationship between classical and quantum geometric systems the corresponding classical and
quantum geometric quadratic algebras are not the same; only the highest order terms in the structure equations agree.

\section{Contractions}

The notion of contractions for quadratic algebras is based on that for Lie algebras, e.g.,~\cite{Wigner, NP, WW}:
\begin{definition}
Let $\mathfrak{g}$ be a~complex Lie algebra with an underlying vector space~$V$ and Lie brackets $ [\;,\;]$.
In the following we simply write it as $\mathfrak{g}=(V,[\;,\;])$.
Suppose that for any $\epsilon\in (0, 1]$, $t_{\epsilon}\colon V\to V$ is a~a linear invertible operator and that
$\lim\limits_{\epsilon\to 0^+}t_\epsilon^{-1}[t_\epsilon X,t_\epsilon Y]$ converges for any $X,Y\in V$.
We use the notation
\begin{gather*}
\lim_{\epsilon\to 0^+}t_\epsilon^{-1}[t_\epsilon X,t_\epsilon Y]=[X,Y]_0.
\end{gather*}
Then $[\;,\;]_0$ are in fact Lie brackets on~$V$ and we denote this Lie algebra by $\mathfrak{g}_0=(V,[\;,\;]_0)$.
We say that $\mathfrak{g}_0$ \textit{is a~contraction of $\mathfrak{g}$} (that is realized by the family of linear maps
$\{t_\epsilon\}_{\epsilon \in (0,1]}$) and we denote it by $\mathfrak{g}\to \mathfrak{g}_0$.
\end{definition}
Thus, as $\epsilon\to 0$ the 1-parameter family of basis transformations can become singular but the structure constants
go to a~f\/inite limit.

\medskip

\noindent
{\bf Note.} In this paper all of the Lie algebra contractions needed are of standard In\"on\"u--Wigner
type~\cite{NP}, but for our theory on contractions of quadratic algebras and second order superintegrable systems on
constant curvature spaces~\cite{KM2014} IW-contractions are not enough.
In most cases we employ a~generalized In\"on\"u--Wigner contraction (Doebner--Melsheimer type)~\cite{Doebner, NP},
but in some specif\/ic cases we are forced to use a~general contraction of Lie algebras in the sense of Saletan~\cite{Saletan}.
In a~paper under preparation we shall demonstrate that all of these generalizations of In\"on\"u--Wigner contractions
are induced by a~well def\/ined family of contractions of the conformal Lie algebra $so(4,{\mathbb C})$ to itself that
follow from limiting processes for $R$-separable coordinate systems for wave equations, introduced by B\^ocher in his
famous 1894 thesis~\cite{Bocher}.
We call these B\^ocher contractions.

Of particular interest to us are contractions that are induced by~$\epsilon$-dependent local analytic coordinate
transformations $x_j(\epsilon, x'_1,\dots,x'_n)$, $j=1,\dots,n$ on a~manifold $\cal M$ such that the Jacobian $\det
(\frac{\partial {\bf x}}{\partial{\bf x'}})\ne 0$ for $\epsilon\in (0,1]$, but the Jacobian is undef\/ined or nonsingular
in the limit as $\epsilon\to 0$.
\begin{definition}
Suppose that $\mathfrak{g}=(V,[\;,\;]_0)$ is a~contraction of $\mathfrak{g}=(V,[\;,\;])$ that is realized by the family
of linear maps $\{t_\epsilon\}_{\epsilon \in (0,1]}$.
Let $\mathcal{M}$ be a~smooth manifold with a~local coordinate system $(x_1,x_2,\dots,x_n)$ and let
$\psi\colon \mathfrak{g}\to C^{\infty}(\mathcal{TM})$ be an embedding of Lie algebras, where
$C^{\infty}(\mathcal{TM})$ is the space of smooth functions on the cotangent bundle (the phase space) of $\mathcal{M}$
which is equipped with its canonical Poisson brackets as Lie brackets.
Suppose that $x_j(\epsilon, x'_1,\dots,x'_n)$, $j=1,\dots,n$ are~$\epsilon$-dependent local analytic coordinate
transformations such that the Jacobian $\det (\frac{\partial {\bf x}}{\partial{\bf x'}})\ne 0$ for $\epsilon\in (0,1]$.
If the limit
\begin{gather*}
\lim_{\epsilon \to 0^+}\psi(t_{\epsilon}(v))(x_1(\epsilon, x'_1,\dots,x'_n),\dots,x_n(\epsilon, x'_1,\dots,x'_n))
\end{gather*}
converges for any $v\in V$ and def\/ines a~Lie algebra homomorphism from $\mathfrak{g}_0$ into a~local expression of
a~certain space of $C^{\infty}(\mathcal{TM}')$ for some smooth manifold $\mathcal{M}'$.
Then we say that the contraction $\mathfrak{g}\to \mathfrak{g}_0$ is implemented by $x_j(\epsilon,
x'_1,\dots,x'_n)$, $j=1,\dots,n$ and call this procedure a~{\it geometric Lie algebra contraction}.
This distinction between abstract contractions of Lie algebras and geometric contractions has been recognized from the
earliest days of the theory, e.g.,~\cite{Wigner}.
\end{definition}

We give some pertinent examples.

\begin{example}
Consider the complex three dimensional Lie algebra ${\cal G}$\textbf{3} def\/ined by basis elements $\{{\cal P}_1,{\cal
P}_2,{\cal D}\}$ that satisfy $[{\cal P}_1,{\cal P}_{2}]=0$, $[{\cal P}_{1},{\cal D}]={\cal P}_{1}$, $[{\cal P}_{2},{\cal
D}]={\cal P}_{2}$.
This algebra admits an abstract In\"on\"u--Wigner contraction
def\/ined by $t_{\epsilon}({\cal P}_{1})=\epsilon {\cal P}_{1}$,
$t_{\epsilon}({\cal P}_{2})=\epsilon {\cal P}_{2}$,
$t_{\epsilon}({\cal D})={\cal D}$.
In this case the contracted Lie algebra, ${\cal G}$\textbf{3}$_0$, coincides with ${\cal G}$\textbf{3}
and $[{\cal P}_1,{\cal P}_{2}]_0=0$, $[{\cal P}_{1},{\cal D}]_0={\cal P}_{1}$, $[{\cal P}_{2},{\cal D}]_0={\cal P}_{2}$.

Now considering the complex analytic manifold $\mathbb{C}^2$ with coordinates $(x_1,x_2)$ we can reali\-ze~${\cal
G}$\textbf{3}~by
\begin{gather*}
\psi({\cal P}_{1})(x_1,x_2,p_{x_1},p_{x_2})= p_{x_1},
\qquad
\psi({\cal P}_2)(x_1,x_2,p_{x_1},p_{x_2})= p_{x_2},
\\
\psi({\cal D})(x_1,x_2,p_{x_1},p_{x_2})= x_1p_{x_1}+x_2p_{x_2}.
\end{gather*}
A~geometric implementation of the contraction is obtained by the substitution $x_1=\epsilon x'_1$, $x_2=\epsilon x'_2$.
Then, taking the limit, we f\/ind
\begin{gather*}
  \lim_{\epsilon \to 0^+} \psi(t_{\epsilon}({\cal P}_{1}))(\epsilon x'_1,\epsilon x'_2,p_{\epsilon
x'_1},p_{\epsilon x'_2})=\lim_{\epsilon \to 0^+} \epsilon p_{\epsilon x'_1}=\lim_{\epsilon \to
0^+} \epsilon \frac{1}{\epsilon}p_{x'_1} =p_{x'_1},
\\
  \lim_{\epsilon \to 0^+} \psi(t_{\epsilon}({\cal P}_{2}))(\epsilon x'_1,\epsilon x'_2,p_{\epsilon
x'_1},p_{\epsilon x'_2})=\lim_{\epsilon \to 0^+} \epsilon p_{\epsilon x'_2}=\lim_{\epsilon \to
0^+} \epsilon \frac{1}{\epsilon}p_{x'_2} =p_{x'_2},
\\
  \lim_{\epsilon \to 0^+} \psi(t_{\epsilon}({\cal D}))(\epsilon x'_1,\epsilon x'_2,p_{\epsilon
x'_1},p_{\epsilon x'_2})=\lim_{\epsilon \to 0^+} \epsilon x'_1 \frac{1}{\epsilon}p_{x'_1}+\epsilon x'_2
\frac{1}{\epsilon}p_{x'_2}=x'_1p_{x'_1}+x'_2p_{x'_2}.
\end{gather*}
Though this Lie algebra contraction acts like the identity map here, we shall see that its action on Darboux quadratic
algebras is nontrivial.
\end{example}
\begin{example}\sloppy
We again consider the Lie algebra ${\cal G}$\textbf{3} with the same geometric identif\/ication.
This algebra admits another abstract In\"on\"u--Wigner contraction
def\/ined by $t_{\epsilon}({\cal P}_{1})={\cal P}_{1}$,
$t_{\epsilon}({\cal P}_{2})={\cal P}_{2}$,
$t_{\epsilon}({\cal D})=\epsilon {\cal D}$.
The contracted Lie algebra, ${\cal G}$\textbf{3}$_0$, is given by $[{\cal P}_1,{\cal P}_{2}]_0=0$, $[{\cal P}_{1},{\cal
D}]_0=0$, $[{\cal P}_{2},{\cal D}]_0=0$.
A~geometric implementation of the contraction is obtained by the substitution $x_1=x'_1+\frac{1}{\epsilon}$, $x_2=x'_2$.
Then, taking the limit, we f\/ind
\begin{gather*}
  \lim_{\epsilon \to 0^+} \psi(t_{\epsilon}({\cal P}_{1}))(x'_1+\frac{1}{\epsilon},
x'_2,p_{x'_1+\frac{1}{\epsilon}},p_{x'_2})=\lim_{\epsilon \to 0^+}
p_{x'_1+\frac{1}{\epsilon}}=\lim_{\epsilon \to 0^+} p_{x'_1} =p_{x'_1},
\\
  \lim_{\epsilon \to 0^+} \psi(t_{\epsilon}({\cal P}_{2}))(x'_1+\frac{1}{\epsilon},
x'_2,p_{x'_1+\frac{1}{\epsilon}},p_{x'_2})=\lim_{\epsilon \to 0^+} p_{x'_2} =p_{x'_2},
\\
  \lim_{\epsilon \to 0^+} \psi(t_{\epsilon}({\cal D}))({\cal P}_{2})
\left(x'_1+\frac{1}{\epsilon}, x'_2,p_{x'_1+\frac{1}{\epsilon}},p_{x'_2}\right)
=\lim_{\epsilon \to 0^+}\epsilon \left(\left(x'_1+\frac{1}{\epsilon}\right)p_{x'_1+\frac{1}{\epsilon}}+ x'_2 p_{x'_2}\right)
 =p_{x'_1}.
\end{gather*}
Hence the resulting map from ${\cal G}$\textbf{3}$_0$ that is given by ${\cal P}_{1}\mapsto p_{x'_1}$, ${\cal
P}_{2}\mapsto p_{x'_2}$, ${\cal D}_{1}\mapsto p_{x'_1}$ is a~Lie algebra homomorphism with kernel spanned~by
${\cal P}_{1}-{\cal D}$.
Though this geometric implementation is not an isomorphism of the contracted Lie algebra we shall see that its action on
Darboux quadratic algebras is isomorphic.
\end{example}
There are exactly analogous implementations of geometric contractions in the quantum case.

\begin{definition}[algebraic contraction of quadratic algebras] Let $\mathcal{A}$ be a~classical nondegenerate quadratic algebra
with a~generating set which consist of a~Hamiltonian, $\mathcal{H}$ (second order element which lies in the center of
$\mathcal{A}$) and two second order constants of motion $\mathcal{L}_1$, $\mathcal{L}_2$.
Let $\{\mathcal{L}_1,\mathcal{L}_2\}^2=\mathcal{R}^2=\mathcal{F}(\mathcal{H},\mathcal{L}_1,\mathcal{L}_2)$ be the
Casimir of $\mathcal{A}$.
For any $\epsilon\in (0,1]$ consider a~matrix $A_{\epsilon}\in {\rm GL}(3,{\mathbb C})$ of the form
\begin{gather*}
A_{\epsilon}=\left(
\begin{matrix} A_{1,1}(\epsilon) & A_{1,2}(\epsilon)&A_{1,3}(\epsilon)
\\
A_{2,1}(\epsilon)&A_{2,2}(\epsilon) &A_{2,3}(\epsilon)
\\
0 &0 &A_{3,3}(\epsilon)
\end{matrix}
\right)\in {\rm GL}(3).
\end{gather*}
Assume that the map from $(0,1]$ to ${\rm GL}(3,{\mathbb C})$ that is given by $\epsilon \mapsto A_{\epsilon}$ is
continuous.
For any $\epsilon\in (0,1]$
we have another set of generators, $\{\mathcal{L}_1^{\epsilon}, \mathcal{L}_2^{\epsilon}, \mathcal{H}^{\epsilon}\}$,
for $\mathcal{A}$ that is def\/ined~by
\begin{gather*}
\left(
\begin{matrix} \mathcal{L}_1
\\
\mathcal{L}_2
\\
\mathcal{H}
\end{matrix}
\right) =A_{\epsilon}\left(
\begin{matrix} \mathcal{L}_1^{\epsilon}
\\
\mathcal{L}_2^{\epsilon}
\\
\mathcal{H}^{\epsilon}
\end{matrix}
\right)
\end{gather*}
and satisf\/ies the Casimir relation
\begin{gather*}
  \{\mathcal{L}^{\epsilon}_1,\mathcal{L}^{\epsilon}_2\}^2=(\mathcal{R}^{\epsilon})^2=\mathcal{R}^2(A_{11}(\epsilon)A_{22}(\epsilon)
-A_{12}(\epsilon)A_{21}(\epsilon))^{-2}
\\
\hphantom{\{\mathcal{L}^{\epsilon}_1,\mathcal{L}^{\epsilon}_2\}^2}{}
=(A_{11}(\epsilon)A_{22}(\epsilon) -A_{12}(\epsilon)A_{21}(\epsilon))^{-2}\mathcal{F}\left(A_{\epsilon}\left(
\begin{matrix}
\mathcal{L}_1^{\epsilon}
\\
\mathcal{L}_2^{\epsilon}
\\
\mathcal{H}^{\epsilon}
\end{matrix}
\right)\right)
\\
\hphantom{\{\mathcal{L}^{\epsilon}_1,\mathcal{L}^{\epsilon}_2\}^2}{}
=\sum\limits_{i+j+k=3,\,0\leq i,j,k}\alpha_{ijk}(\epsilon)(\mathcal{L}_1^{\epsilon})^i(\mathcal{L}_2^{\epsilon})^j
(\mathcal{H}^{\epsilon})^k.
\end{gather*}
If $\lim\limits_{\epsilon \to 0^+}\alpha_{ijk}(\epsilon)$ exists for any $i$, $j$, $k$ we denote it by $\alpha_{ijk}(0)$.
Then there exists a~quadratic algebra, $\mathcal{A}_0$ with a~set of generators $\{\mathcal{L}_1^{0}, \mathcal{L}_2^{0}, \mathcal{H}^{0}\}$
that satisfy
\begin{gather*}
\big\{\mathcal{L}^{0}_1,\mathcal{L}^{0}_2\big\}^2=\big(\mathcal{R}^0\big)^2=\sum\limits_{i+j+k=3,\,0\leq
i,j,k}\alpha_{ijk}(0)\big(\mathcal{L}_1^{0}\big)^i\big(\mathcal{L}_2^{0}\big)^j \big(\mathcal{H}^{0}\big)^k
\end{gather*}
we call $\mathcal{A}_0$ the contraction of $\mathcal{A}$ with respect to $\{A_{\epsilon}\}_{\epsilon \in (0,1]}$.
\end{definition}

Note that we can expand $\{{\mathcal R}^\epsilon,{\mathcal L}_1^\epsilon\}$, $\{{\mathcal R}^\epsilon,{\mathcal
L}_2^\epsilon\}$ as quadratic expressions in ${\mathcal L}_1^\epsilon$, ${\mathcal L}_2^\epsilon$, $\mathcal{H}^{\epsilon}$.
For a~contraction it might seem that we must also require these expansion coef\/f\/icients to have f\/inite limits as
$\epsilon\to 0$.
However from the results of Section~\ref{structure} the convergence of these other structure equations follows from the
convergence of the Casimir.
There is a~completely analogous def\/inition of contraction for quantum quadratic algebras.

Just as for abstract classical and quantum Lie algebra contractions there are abstract classical and quantum quadratic
algebra contractions that are induced by~$\epsilon$-dependent local analytic coordinate transformations $x_j(\epsilon,
x'_1,\dots,x'_n)$, $j=1,\dots,n$ on a~manifold $\cal M$ such that the Jacobian $\det (\frac{\partial {\bf
x}}{\partial{\bf x'}})\ne 0$ for $\epsilon\in (0,1]$, but the Jacobian is undef\/ined or nonsingular in the limit as
$\epsilon\to 0$.
If an algebraic contraction $A\to B$ can be implemented by some coordinate transformation, we say that it is
a~{\it geometric quadratic algebra contraction}.
In fact, all of the quadratic algebra contractions for Darboux systems discussed in this paper are geometric
implementations.
We will give many examples in the following sections.

The notion of contraction applied to structures other than Lie algebras is not new, see for example~\cite{Gromov} (and
references there-in) and~\cite{Mad}.

In~\cite{KM2014} Lie algebra and quadratic algebra contractions for superintegrable systems on constant curvature spaces
were related:
\begin{theorem}\label{theorem 4}
Every Lie algebra contraction of ${\cal G}=e(2,{\mathbb C})$ or ${\cal G}=o(3,{\mathbb C})$ induces a~geometric
contraction of a~free geometric quadratic algebra $\tilde Q$ based on $\cal G$, which in turn induces uniquely
a~contraction of the quadratic algebra~$Q$ with potential.
This is true for both classical and quantum algebras.
\end{theorem}

Here we will demonstrate the analogous result for Darboux spaces, using the conformal symmetry algebra ${\cal
G}3$ with basis $\{\partial_x,\partial_y, x\partial_x+y\partial_y\}$.

\section{Structure relations}

Although the full sets of classical structure equations can be rather complicated, the func\-tion~$\cal F$ contains all of
the structure information for nondegenerate systems and $\cal G$ (only unique up to a~nonzero scalar multiple) most of
the information for degenerate systems.
In particular, it is easy to show that~\cite{Perelomov, KM2014},
$\{{\cal L}_1,{\cal R}\}=\frac12\frac{\partial F}{\partial {\cal L}_2}$,
$\{{\cal L}_2,{\cal R}\}=-\frac12\frac{\partial F}{\partial {\cal L}_1}$,
so the Casimir contains within itself all of the structure equations.
For degenerate systems we have~\cite{KM2014}
\begin{gather*}
\{{\cal X},{\cal L}_1\}=K\frac{\partial {\cal G}}{\partial {\cal L}_2},
\qquad
\{{\cal X},{\cal L}_2\}=-K\frac{\partial{\cal G}} {\partial {\cal L}_1},
\qquad
\{{\cal L}_1,{\cal L}_2\}=K\frac{\partial{\cal G}}{\partial {\cal X}},
\end{gather*}
where $\{{\cal H},K\}=0$.
Here,~$K$ is a~scalar, unless $\{{\cal X},{\cal L}_1\}$ and $\{{\cal X},{\cal L}_2\}$ are linearly dependent.
In the latter case there would exist 3 algebraically independent elements of the algebra in involution, including the
Hamiltonian.
This is impossible for a~Hamiltonian system.
Thus, except for some abstract quadratic algebras unrelated to geometric superintegrable systems,~$K$ will always be
a~scalar that can be normalized to~1.

\subsection[The quantum operators $F$ and $G$]{The quantum operators $\boldsymbol{F}$ and $\boldsymbol{G}$}
\label{structure}

The quantum case is similar to the classical case, but more complicated.
From the Casimir relation
\begin{gather*}
R^2-F\equiv R^2 - \big(b_1 L_1^3 + b_2 L_2^3 + b_3 H^3 + b_4 \big\{L_1^2,L_2\big\} + b_5 \big\{L_1,L_2^2\big\} + b_6 L_1 L_2 L_1 + b_7 L_2 L_1 L_2
\\
\hphantom{R^2-F\equiv}{}
{}+ b_8 H\{L_1,L_2\} +b_9 H L_1^2 + b_{10} H L_2^2 + b_{11} H^2 L_1 + b_{12} H^2 L_2 + b_{13} L_1^2 + b_{14} L_2^2
\\
\hphantom{R^2-F\equiv}{}
{}+ b_{15} \{L_1,L_2\}
+ b_{16} H L_1 + b_{17} H L_2 + b_{18} H^2 + b_{19} L_1 + b_{20} L_2 + b_{21} H + b_{22}\big) = 0,
\end{gather*}
we want to determine the structure relations.
Noting that $R=[L_1,L_2]$ and using operator identities
\begin{gather}
L_2RL_2 = \tfrac12[[L_2,R],L_2]+\tfrac12\big\{R,L_2^2\big\},
\qquad
L_1RL_1=\tfrac12[[L_1,R],L_1]+\tfrac12\big\{R,L_1^2\big\},
\nonumber
\\
L_1RL_2+L_2RL_1 = -\tfrac12[L_2,[L_1,R]]-\tfrac12[L_1,[L_2,R]]+\tfrac12\{R,\{L_1,L_2\}\},
\label{opident}
\\
{}\big[L_1,R^2\big] = \{R,[L_1,R]\},
\qquad
\big[L_2,R^2\big]=\{R,[L_2,R]\},
\nonumber
\\
{}[L_1,L_2L_1L_2] = \big\{[L_1,L_2],\tfrac12\{L_1,L_2\}\big\},
\nonumber
\\
XL_1L_2+L_2L_1X=\tfrac12\{X,\{L_1,L_2\}\}+\tfrac12[X,[L_1,L_2]],
\nonumber
\end{gather}
(true formally for all operators $L_1$, $L_2$, $R$, not just for~$R$ the commutator), and setting
\begin{gather*}
[L_1,R]=A_1L_1^2+A_2L_2^2+A_3H^2+A_4\{L_1,L_2\}
\\
\phantom{[L_1,R]=}{}
+A_5HL_1+A_6HL_2+A_7L_1+A_8L_2+A_9H+A_{10},
\\
[L_2,R]=B_1L_1^2+B_2L_2^2+B_3H^2+B_4\{L_1,L_2\}
\\
\phantom{[L_1,R]=}{}
+B_5HL_1+B_6HL_2+B_7L_1+B_8L_2+B_9H+B_{10},
\end{gather*}
we can write $[L_j,R^2-F]=0 $ in the form $\{R,Q_j\}=0$ for some explicit dif\/ferential opera\-tor~$Q_j$.
This can only hold if $Q_j=0$.
As a~result we f\/ind:
\begin{gather*}
A_1 = b_4 + \tfrac12 b_6,
\qquad
A_2 = \tfrac32 b_2,
\qquad
A_3 = \tfrac12 b_{12},
\qquad
A_4 = b_5 + \tfrac12 b_7,
\\
A_5 = b_8,\qquad A_6 = b_{10}, \qquad
A_7 = b_{15} - \tfrac34 b_1 b_2 + b_4 b_5 -\tfrac14 b_6 b_7,
\\
A_8 = b_{14} -\tfrac12 b_2 b_4 + b_5^2 - b_2 b_6 + \tfrac12 b_5 b_7, \qquad A_9 = \tfrac12 b_{17} - \tfrac14 b_2 b_9 + \tfrac12 b_5 b_8 -\tfrac14 b_6 b_{10},
\\
A_{10} = \tfrac12 b_{20} - \tfrac14 b_2b_{13} -\tfrac14 b_1 b_2 b_5
\\
\phantom{A_{10}=}{}
-\tfrac14 b_2 b_4^2 +\tfrac12 b_5 b_{15} +\tfrac12 b_4 b_5^2 +\tfrac14 b_1 b_2 b_7 + \tfrac14 b_2 b_6^2 -\tfrac14 b_5^2 b_6 -
\tfrac14 b_5 b_6 b_7 -\tfrac14 b_6 b_{14},
\\
B_1 = -\tfrac32 b_1,
\qquad
B_2 = -b_5 -\tfrac12 b_7,
\qquad
B_3 = -\tfrac12 b_{11},
\qquad
B_4 = -b_4 - \tfrac12 b_6,
\\
B_5 = -b_9,
\qquad
B_6 = -b_8,
\qquad
B_7 = -b_{13} +\tfrac12 b_1 b_5 - b_4^2 + b_1 b_7 - \tfrac12 b_4 b_6,
\\
B_8 = -b_{15} + \tfrac34 b_1 b_2 - b_4 b_5 + \tfrac14 b_6 b_7,
\qquad
B_9 = -\tfrac12 b_{16} + \tfrac14 b_1 b_{10} - \tfrac12 b_4 b_8 + \tfrac14 b_7 b_9,
\\
B_{10}= -\tfrac12 b_{19} + \tfrac14 b_1b_{14} +\tfrac14 b_1 b_2 b_4 + \tfrac14 b_1 b_5^2 - \tfrac12 b_4 b_{15}
\\
\phantom{B_{10}=}{}
- \tfrac12 b_4^2 b_5 - \tfrac14 b_1 b_2 b_6 -\tfrac14 b_1 b_7^2 +\tfrac14 b_4^2 b_7 + \tfrac14 b_4 b_6 b_7 + \tfrac14 b_7 b_{13}.
\end{gather*}

For quantum degenerate systems, in the Casimir relation
\begin{gather*}
G(L_1,L_2,H,X, \alpha)=0
\end{gather*}
for 2nd order superintegrable systems with degenerate potentials we assume that~$G$ is given, up to a~multiplicative
factor, and set
\begin{gather*}
G=c_1 L_1^2+c_2L_2^2+c_3H^2+c_4\{L_1,L_2\}+c_5HL_1+c_6HL_2+c_7X^4+c_8\big\{X^2,L_1\big\}
\\
\phantom{G=}{}+c_9\big\{X^2,L_2\big\}
+c_{10}HX^2+c_{11}XL_1X+c_{12}XL_2X+c_{13}L_1+c_{14}L_2\\
\phantom{G=}{}
+c_{15}H+c_{16}X^2+c_{17},
\\
[X,L_1]=C_1L_1+C_2L_2+C_3H+C_4X^2+C_5,
\\
[X,L_2]=D_1L_1+D_2L_2+D_3H+D_4X^2+D_5,
\\
[L_1,L_2]=E_1\{L_1,X\} +E_2\{L_2,X\} +E_3HX+E_4X^3+E_5X.
\end{gather*}
Using identities~\eqref{opident} we f\/ind
\begin{gather}
[X,G]= \Big\{[X,L_1],A_1-\frac{c_{11}}{4}\big(C_{1}^2 + C_2 D_1\big) -\frac{c_{12}}{4}(C_1 D_1 + D_1 D_2) + \frac{c_{13}}{2}\Big\}
\nonumber
\\
\label{GX}
\phantom{[X,G]=}{}
 + \Big\{[X,L_2],A_2+\frac{c_{11}}{4}(C_1 C_2 + C_2 D_2) - \frac{c_{12}}{4}(C_2 D_1 + D_2^2) + \frac{c_{14}}{2}\Big\}=0,
\\
[L_1,G] = \Big\{[L_1,L_2],A_2+\frac{c_8}{4}(C_1 C_2 + C_2 D_2) - \frac{c_9}{4}\big(C_1^2 + C_1 D_2\big) + \frac{c_{12}}{4}
(C_1 D_2 - C_2 D_1)
\nonumber
\\
\label{GS1}
\phantom{[L_1,G]=}{}
 + \frac{c_{14}}{2} \Big\}
 + \{[L_1,X],A_3 +k_1 X \} + \{[L_2,X], k_2 X\}=0,
\\
[L_2,G] = \Big\{[L_1,L_2],-A_1+\frac{c_8}{4}\big(C_1 D_2 + D_2^2\big) -\frac{c_9}{4}(C_1 D_1 + D_1 D_2) + \frac{c_{11}}{4} (C_2
D_1 - C_1 D_2)
\nonumber
\\
\label{GS2}
\phantom{[L_2,G]=}{}
-\frac{c_{13}}{2}\Big\}
 + \{[L_1,X],k_3 X\} + \{[L_2,X], A_3 +k_4 X\}=0,
\end{gather}
where
\begin{gather*}
A_1=c_1L_1+c_4L_2+\frac{c_5}{2}H+\left(c_8+\frac{c_{11}}{2}\right)X^2,
\\
A_2=c_2L_2+c_4L_1+\frac{c_6}{2}H+\left(c_9+\frac{c_{12}}{2}\right)X^2,
\\
A_3=2c_7X^3+\left(c_8+\frac{c_{11}}{2}\right)\{L_1,X\}+\left(c_9+\frac{c_{12}}{2}\right)\{L_2,X\}+c_{10}HX,
\\
k_1= -c_7 (C_1^2 + C_2 D_1) + \frac{c_8}{2}(2 C_1 C_4 + C_2 D_4 - C_2 E_1) + \frac{c_9}{2} (2 C_4 D_1 + C_1 E_1 - C_1 D_4)
\\
\phantom{k_1=}{}
+ \frac{c_{12}}{2} (C_1 E_1 + C_1 D_4 + D_1 E_2 - C_4 D_1)+c_{16},
\\
k_2 = -c_7(C_1 C_2 + C_2 D_2) + \frac{c_8}{2}(C_2 C_4 - C_2 E_2) + \frac{c_9}{2} (2 C_4 D_2 +C_1 C_4 + C_1 E_2)
\\
\phantom{k_2=}{}
+ \frac{c_{12}}{2} (C_2 D_4 + C_2 E_1 + D_2 E_2 - C_4 D_2),
\\
k_3 = -c_7(C_1 D_1 +D_1 D_2) + \frac{c_8}{2}(2 C_1 D_4 + D_2 D_4 - D_1 E_1) + \frac{c_9}{2}(D_1 E_1 + D_1 D_4)
\\
\phantom{k_3=}{}
+ \frac{c_{11}}{2} (C_4 D_1 - C_1 D_4 -C_1 E_1 -D_1 E_2),
\\
k_4 = -c_7 (C_2 D_1 + D_2^2) + \frac{c_8}{2}(2 C_2 D_4 - C_4 D_2 - D_2 E_2) +\frac{c_9}{2} (2 D_2 D_4 + C_4 D_1 + D_1 E_2)
\\
\phantom{k_4=}{}
+ \frac{c_{11}}{2} (C_4 D_2 - C_2 D_4 - C_2 E_1 - D_2 E_2) + c_{16}.
\end{gather*}
Equating the coef\/f\/icients of the 4th order terms in~\eqref{GX} and the coef\/f\/icients of the 5th order terms
in~\eqref{GS1} and~\eqref{GS2} we f\/ind
\begin{gather*}
[X,L_1]=KA_2+C_5,
\qquad
[X,L_2]=-KA_1+D_5,
\qquad
[L_1,L_2]=KA_3+E_5X,
\\
C_1 = Kc_4,
\qquad
C_2=Kc_2,
\qquad
C_3=K\frac{c_6}{2},
\qquad
C_4=K\left(c_9+\frac{c_{12}}{2}\right),
\\
D_1 = -Kc_1,
\qquad
D_2=-Kc_4,
\qquad
D_3=-K\frac{c_5}{2},
\qquad
D_4=-K\left(c_8+\frac{c_{11}}{2}\right),
\\
E_1 = K\left(c_8+\frac{c_{11}}{2}\right),
\qquad
E_2=K\left(c_9+\frac{c_{12}}{2}\right),
\qquad
E_3=Kc_{10},
\qquad
E_4=2Kc_7,
\end{gather*}
for some constant~$K$.
Now we substitute these values back into~\eqref{GX},~\eqref{GS1}, and~\eqref{GS2}.
We immediately see that $k_2=k_3=0$ and
\begin{gather*}
k_1= k_4=c_{16} -K^2 c_7 \big(c_4^2 -c_1 c_2\big) +K^2 (c_4 c_9 - c_2 c_8)\left(c_8 + \frac{c_{11}}{2}\right)\\
\hphantom{k_1= k_4=}{}  + K^2 (c_4 c_8 - c_1 c_9)
\left(c_9 + \frac{c_{12}}{2}\right)
\end{gather*}
and we obtain
\begin{gather}
C_5A_1+D_5A_2+\left(\frac{c_{13}}{2}-K^2 \frac{c_{11}}{4}\big(c_4^2-c_1 c_2\big)\right) (K A_2 + C_5)
\nonumber
\\
\qquad
{}+\left(\frac{c_{14}}{2}-K^2 \frac{c_{12}}{4}\big(c_4^2-c_1 c_2\big)\right)(-KA_1+D_5) = 0,
\label{GXa}
\\
2 \left(\frac{c_{14}}{2}-K^2 \frac{c_{12}}{4}\big(c_4^2-c_1 c_2\big)\right) (K A_3 + E_5 X) - 2 C_5 (A_3 + k_1 X)
\nonumber
\\
\qquad
{}+(E_5 - Kk_1) \{X,A_2\} = 0,
\label{GS1a}
\\
-2\left(\frac{c_{13}}{2}-K^2 \frac{c_{11}}{4}\big(c_4^2-c_1 c_2\big)\right) (KA_3 + E_5 X) - 2 D_5 (A_3 + k_1 X)
\nonumber
\\
\qquad
{}-(E_5-K k_1) \{X,A_1\} = 0.
\label{GS2a}
\end{gather}

These equations split into terms of order 3,2,1 and 0.
From equation~\eqref{GXa} we f\/ind
\begin{gather}
\label{C5D5}
C_5=\frac{c_{14}}{2}K-\frac{c_{12}}{4}K^3\big(c_4^2-c_1c_2\big),
\qquad
D_5=-\frac{c_{13}}{2}K+\frac{c_{11}}{4}K^3\big(c_4^2-c_1c_2\big)
\end{gather}
except, possibly, for some degenerate cases.
The condition that~\eqref{C5D5} is the unique solution of~\eqref{GXa} is exactly that the set $([X,S_1], [X,S_2])$ is
linearly independent.
Otherwise the solution, though it always exists, is not unique.

Substituting~\eqref{C5D5} into~\eqref{GS1a} and~\eqref{GS2a}, we have
\begin{gather*}
\frac{2 C_5 (E_5-K k_1)}{K} X + (E_5 - K k_1) \{X, A_2\}  = 0,
\\
\frac{2 D_5 (E_5 - K k_1)}{K} X - (E_5 - K k_1) \{X, A_1\} =0,
\end{gather*}
whence we f\/ind
\begin{gather*}
E_5 =K c_{16} -K^3 c_7 \big(c_4^2 -c_1 c_2\big) +K^3 (c_4 c_9 - c_2 c_8)\left(c_8 + \frac{c_{11}}{2}\right)\\
\hphantom{E_5 =}{}  + K^3 (c_4 c_8 - c_1 c_9)
\left(c_9+\frac{c_{12}}{2}\right).
\end{gather*}

We conclude in both the classical and quantum cases that the Casimirs of superintegrable systems determine the structure
equations.

\section{Free 2D 2nd order superintegrable systems}

As was shown in~\cite{KKM20041,KKM20041+1,KKM20041+2,KKM20041+3} the ``free'' 2nd order superintegrable system obtained
by setting all the parameters in a~nondegenerate potential equal to zero retains all of the information needed to
reconstruct the potential.
Thus we can, in principle, restrict our attention to free systems.
First we review from~\cite{KKM20041,KKM20041+1,KKM20041+2,KKM20041+3,KM2014} how the structure equations for 2D 2nd
order nondegenerate classical superintegrable systems are determined.
Such a~system admits a~symmetry ${\cal L} =\sum a^{ij}p_ip_j+W$ if and only if $\{{\cal H},{\cal L}\}=0$, i.e., the
Killing equations are satisf\/ied and $W_i=\lambda\sum\limits_{j=1}^2 a^{ij}V_j$.
Here $W_i=\partial_{x_1}W$ with a~similar convention for subscripts on~$V$.
The equations for~$W$ can be solved provided the Bertrand--Darboux equation $\partial_{x_1}W_{2}=\partial_{x_2}W_1$
holds.
For a~superintegrable system with independent symmetries ${\cal L}_1 =\sum a^{ij}p_ip_j+W^{(1)}$, ${\cal L}_2 =\sum
b^{ij}p_ip_j+W^{(2)}$, we can solve the two independent Bertrand--Darboux equations for the potential to obtain the
canonical system
\begin{gather}
\label{canonicalequations}
V_{22}-V_{11}=A^{22}V_1+B^{22}V_2,
\qquad
V_{12}=A^{12}V_1+B^{12}V_2.
\end{gather}
Here,
\begin{gather*}
A^{12}=-G_2+\frac{D_{(2)}}{D},
\qquad
A^{22}=2G_1+\frac{D_{(3)}}{D},
\\
B^{12}=-G_1-\frac{D_{(0)}}{D},
\qquad
B^{22}=-2G_2-\frac{D_{(1)}}{D},
\\
D=\det \left(
\begin{matrix} a^{11}-a^{22}& a^{12}
\\
b^{11}-b^{22}& b^{12}
\end{matrix}
\right),
\qquad
D_{(0)}=\det \left(
\begin{matrix} 3a^{12}_2& -a^{12}
\\
3b^{12}_2& -b^{12}
\end{matrix}
\right),
\\
D_{(1)}=\det \left(
\begin{matrix} 3a^{12}_2& a^{11}-a^{22}
\\
3b^{12}_2& b^{11}-b^{22}
\end{matrix}
\right),
\qquad
D_{(2)}=\det \left(
\begin{matrix} 3a^{12}_1& a^{12}
\\
3b^{12}_1& b^{12}
\end{matrix}
\right),
\\
D_{(3)}=\det \left(
\begin{matrix} 3a^{12}_1& a^{11}-a^{22}
\\
3b^{12}_1& b^{11}-b^{22}
\end{matrix}
\right),
\end{gather*}
where $\lambda=\exp G$.
If the integrability equations for~\eqref{canonicalequations} are satisf\/ied identically then the solution space is
4-dimensional and we can always express the solution in the form $V({\bf x})= \sum\limits_{j=1}^3a_jV_{(j)}({\bf x})+a_4$,
where $a_4$ is a~trivial additive constant.
In this case the potential is {\it nondegenerate} and 3-parameter.
Another possibility is that the solution space is 2-dimensional with general solution $V({\bf x})= a_1V_{(1)}({\bf x})+a_2$.
For nondegenerate superintegrability, the integrability conditions for the canonical equations must be satisf\/ied
identically, so that $V$, $V_1$, $V_2$, $V_{11}$ can be prescribed arbitrarily at a~f\/ixed regular point.

To obtain the integrability conditions for equations~\eqref{canonicalequations} we introduce the dependent variables
$W^{(1)}=V_1$, $W^{(2)}=V_2$, $W^{(3)}=V_{11}$, and matrices
\begin{gather*}
{\bf w}=\left(
\begin{matrix} W^{(1)}
\\
W^{(2)}
\\
W^{(3)}
\end{matrix}
\right),
\qquad
{\bf A}^{(1)}=\left(
\begin{matrix} 0&0&1
\\
A^{12}&B^{12}&0
\\
A^{13}&B^{13}&B^{12}-A^{22}
\end{matrix}
\right),
\qquad
{\bf A}^{(2)}=\left(
\begin{matrix} A^{12}&B^{12}&0
\\
A^{22}&B^{22}&1
\\
A^{23}&B^{23}&A^{12}
\end{matrix}
\right),
\\
A^{13} = A^{12}_2-A^{22}_1+B^{12}A^{22}+A^{12}A^{12}-B^{22}A^{12},
\\
B^{13} = B^{12}_2-B^{22}_1+A^{12}B^{12},
\qquad
A^{23}= A^{12}_1+B^{12}A^{12},
\qquad
B^{23}=B^{12}_1+B^{12}B^{12}.
\end{gather*}
Then the integrability conditions for system $\partial_{x_j}{\bf w}={\bf A}^{(j)}{\bf w}$, $j=1,2$, must hold:
\begin{gather}
\label{2int3}
A^{(2)}_1-A^{(1)}_2=A^{(1)}A^{(2)}-A^{(2)}A^{(1)}\equiv \big[A^{(1)},A^{(2)}\big].
\end{gather}
If and only if~\eqref{2int3} holds, the system has a~4D vector space of solutions~$V$.

There is a~similar analysis for a~``free'' 2nd order superintegrable system obtained by setting the parameter in a~{\it
degenerate} potential equal to zero,~\cite{KKMP2009}: The free system retains all of the information needed to
reconstruct the potential.
All such degenerate superintegrable systems with potential are restrictions of nondegenerate systems obtained~by
restricting the parameters so that one 2nd order symmetry becomes a~perfect square, e.g.,~${\cal L}_1={\cal X}^2$.
Then ${\cal X}$ is a~1st order constant, necessarily of the form ${\cal X}=\xi_1p_1+\xi_2 p_2$, without a~function term.
Since the degenerate systems are obtained by restriction, the potential function must satisfy the
equations~\eqref{canonicalequations} inherited from the nondegenerate system, with the same functions $A^{ij}$, $B^{ij}$.
In addition the relation $\{{\cal X},{\cal H}\}=0$ imposes the condition $\xi_1V_1+\xi_2V_2=0$.
By relabeling the coordinates, we can always assume $\xi_2\ne 0$ and write the system of equations for the potential in
the form $V_2 = C^2V_1$, $V_{22}=V_{11} + C^{22}V_1$, $V_{12} = C^{12}V_1$, where
\begin{gather*}
C^2(x_1,x_2)=-\frac{\xi_1}{\xi_2},
\qquad
C^{22}(x_1,x_2)=A^{22} -\frac{\xi_1}{\xi_2}B^{22},
\qquad
C^{12}(x_1,x_2)=
A^{12}-\frac{\xi_1}{\xi_2}B^{12}.
\end{gather*}
To f\/ind integrability conditions for these equations we introduce matrices
\begin{gather*}
{\bf v}=\left(
\begin{matrix}
V
\\
V_1
\end{matrix}
\right),
\qquad
{\bf B}^{(1)}=\left(
\begin{matrix}
0&1
\\
0&\partial_2 C^2+C^2C^{12}-C^{22}
\end{matrix}
\right),
\qquad
{\bf B}^{(2)}=\left(
\begin{matrix}
0&C^2
\\
0&C^{12}
\end{matrix}
\right).
\end{gather*}
Then integrability conditions for system $\partial_{x_j}{\bf v}={\bf B}^{(j)}{\bf v}$, $j=1,2$, must hold:
\begin{gather}
\label{2int9}
B^{(2)}_1-B^{(1)}_2=B^{(1)}B^{(2)}-B^{(2)}B^{(1)}\equiv \big[B^{(1)},B^{(2)}\big].
\end{gather}
If and only if~\eqref{2int9} holds, the system has a~2D space of solutions~$V$.
Since $V=\ {\rm constant}$ is always a~solution,
\eqref{2int9} is necessary and suf\/f\/icient for the existence of
a~nonzero 1-parameter potential system.
In this case we can prescribe the values~$V$, $V_2$ at any regular point ${\bf x}_0$; there will exist a~unique $V({\bf x})$ taking these values.

\subsection{Free triplets}

Here we review information about free triplets that was presented and proved in~\cite{KM2014}.
A~{\it $2$nd order classical free triplet} is a~2D system without potential, ${\cal H}_0=\frac{p_1^2+p_2^2}{\lambda(x,y)}$
and with a~basis of~3 functionally independent second-order constants of the motion ${\cal
L}_{(s)}=\sum\limits_{i,j=1}^2a^{ij}_{(s)} p_ip_j$, $a^{ij}_{(s)}=a^{ji}_{(s)}$, $s=1,2,3$, ${\cal L}_{(3)}={\cal H}_0$.
Since the duals of these constants of the motion are 2nd order Killing tensors, the spaces associated with free triplets
can be characterized as 2D manifolds that admit 3 functionally independent 2nd order Killing tensors.
As mentioned above, they have been classif\/ied in~\cite{Koenigs}.
Since the vectors $\{{\bf h_{(s)}}\}$, ${\bf h_{(s)}}^{\rm tr}(x,y,z)=(a^{11}_{(s)}, a^{12}_{(s)},
a^{22}_{(s)})$ form a~linearly independent set, there exist unique $3\times 3$ matrices ${\cal C}^{(j)}$ such that
$ \partial_{x_j}{\bf h}_{(s)}={\cal C}^{(j)}{\bf h}_{(s)}$, $j,s=1,2$.
By linearity, any element ${\cal L}=\sum\limits_{i,j=1}^2a^{ij} p_ip_j$ of the space of 2nd order symmetries spanned~by
the basis triplet is characterized by matrix equations
\begin{gather}
\label{2int6}
\partial_{x_j}{\bf h}={\cal C}^{(j)}{\bf h},
\qquad
j=1,2,
\qquad
{\bf h}^{\rm tr}(x,y,z)=\left(a^{11},a^{12},a^{22}\right).
\end{gather}
In particular, at any regular point ${\bf x}_0$ we can arbitrarily choose the value of the 3-vector ${\bf h}_0$ and
solve~\eqref{2int6} to f\/ind the unique symmetry $\cal L$ of ${\cal H}_0$ such that ${\bf h}({\bf x}_0)={\bf h}_0$.
A~normalization condition for the ${\cal C}^{(j)}$:~\eqref{2int6} is valid for $a^{11}=a^{22}={1}/{\lambda}$, $a^{12}=0$,
i.e., for ${\cal H}_0$.
From this and the requirement that the $\cal L$ are free constants of the motion we f\/ind
\begin{gather*}
{\cal C}^{(1)}=\left(
\begin{matrix}
-G_1&-G_2&0
\vspace{1mm}\\
-\frac12{\cal C}^{(2)}_{11}&-\frac12G_1-\frac12{\cal C}^{(2)}_{12}&\frac12{\cal C}^{(2)}_{11}
\vspace{1mm} \\
-G_1-2{\cal C}^{(2)}_{21}& -G_2-2{\cal C}^{(2)}_{22}&2{\cal C}^{(2)}_{21}
\end{matrix}
\right),
\qquad
{\cal C}^{(2)}=\left(
\begin{matrix}
{\cal C}^{(2)}_{11}&{\cal C}^{(2)}_{12}&-G_2-{\cal C}^{(2)}_{11}
\vspace{1mm}\\
{\cal C}^{(2)}_{21}&{\cal C}^{(2)}_{22}&-{\cal C}^{(2)}_{21}
\vspace{1mm}\\
0 & -G_1 &-G_2
\end{matrix}
\right),
\end{gather*}
with the 4 functions ${\cal C}^{(2)}_{11}$, ${\cal C}^{(2)}_{12}$, ${\cal C}^{(2)}_{21}$, ${\cal C}^{(2)}_{22}$ free.
If we def\/ine the functions $A^{12}$, $B^{12}$, $A^{22}$, $B^{22}$ by the requirement
\begin{gather*}
{\cal C}^{(2)}_{11}=-\frac23 G_2-\frac23 A^{12},
\qquad
{\cal C}^{(2)}_{12}=\frac13 G_1-\frac23 A^{22},
\\
{\cal C}^{(2)}_{21}=-\frac13 G_1-\frac13B^{12},
\qquad
{\cal C}^{(2)}_{22}=-\frac23G_2-\frac13 B^{22},
\end{gather*}
then equations~\eqref{2int6} agree with the equations that are obtained from a~superintegrable system with nondegenerate
potential satisfying~\eqref{canonicalequations}.
Thus, for a~free system there always exist unique functions $A^{ij}$, $B^{ij}$.
Then necessary and suf\/f\/icient conditions for extension to a~system with nondegenerate potential~$V$ satisfying
equations~\eqref{canonicalequations} are that conditions~\eqref{2int3} hold identically.

This analysis also extends, via restriction, to superintegrable systems with degenerate potential.
A~free triplet that corresponds to a~degenerate superintegrable system is one that corresponds to a~nondegenerate system
but such that one of the free generators can be chosen as a~perfect square.
For these systems conditions~\eqref{2int9} for the potential are satisf\/ied identically.

Similarly, we def\/ine a~{\it $2$nd order quantum free triplet} as a~2D quantum system without potential,
$H_0=\frac{1}{\lambda({\bf x})}(\partial_{11}+\partial_{22})$, and with a~basis of 3 algebraically independent
second-order symmetry operators
\begin{gather*}
L_k=\frac{1}{\lambda}\sum\limits_{i,j=1}^2\partial_{i}(\lambda a^{ij}_{(k)}\partial_j) ({\bf x}),
\qquad
k=1,2,3,
\qquad
a^{ij}_{(k)}=a^{ji}_{(k)},
\qquad
L_3=H_0.
\end{gather*}
There is a~1-1 relationship between classical and quantum free triplets.

\section{Free Darboux systems}

The Darboux spaces admit a~1-dimensional space of Killing vectors and a~4-dimensional space of 2nd order Killing tensors.
Thus each space can admit at most one superintegrable system with degenerate potential, and each space does so.
We merely need to check that equations~\eqref{2int3} are satisf\/ied.
Then we can compute the degenerate potential.
Turning of\/f the 1-parameter potential produces a~single free degenerate quadratic algebra which we list below.
There are no more possibilities.
There are a~number of possibilities for free triplets to def\/ine a~nondegenerate quadratic algebra for a~Darboux space,
however.
We classify the possibilities up to conjugacy under the action of the 1-parameter symmetry group of the manifold.
Note that the 4-dimensional space of free constants of the motion is not obtained from the enveloping algebra of an
underlying symmetry group.
We shall see that there is a~1-1 relationship between free quadratic algebras and restrictions of quadratic algebras of
nondegenerate superintegrable systems.
We adopt the labeling of superintegrable systems and constants of the motion on Darboux spaces introduced
in~\cite{KKMW}, using a~tilde to dif\/ferentiate between a~free triplet and its associated superintegrable system.
In the following sections, with the aid of MAPLE, we classify all possible free quadratic algebras generated by the 2nd
order Killing tensors, up to conjugacy.
Then, using MAPLE again, we verify for each quadratic algebra that the integrability conditions~\eqref{2int9} are
satisf\/ied and we compute the nondegenerate potentials.
Most of the results are presented in lists but in Section~\ref{example1}
we give more details on the construction of the
superintegrable system with potential in one case.

Each of the Darboux spaces can be embedded as a~surface in 3 dimensions if we regard the ignorable variable as an angle,
i.e.,~$X=f(x)\cos(y)$, $Y=f(x)\sin(y)$, $Z=h(x)$, and this is not unique~\cite{Eisenhart66}.
We give an illustrative example for each case.

\subsubsection{Free Darboux 1 systems}

The space Darboux 1 ($D_1$) has free degenerate Hamiltonian
\begin{gather*}
{\bf \tilde D1D}\colon \quad
{\cal H}=\frac{1}{4x}\big(p_x^2+p_y^2\big)
\end{gather*}
with a~single Killing vector ${\cal K}=p_y$ and a~basis, $\{{\cal H}, {\cal K}^2, {\cal X}_1, {\cal X}_2\}$ for the
4-dimensional space of 2nd order Killing tensors.
Here,
\begin{gather*}
{\cal X}_1=p_xp_y-\frac{y}{2x}\big(p_x^2+p_y^2\big),
\qquad
{\cal X}_2=p_y(yp_x-xp_y)-\frac{y^2}{4x}\big(p_x^2+p_y^2\big).
\end{gather*}
The commutation relations are
\begin{gather*}
\{{\cal K},{\cal X}_1\}=-2{\cal H},
\qquad
\{{\cal K},{\cal X}_2\}={\cal X}_1,
\qquad
\{{\cal X}_1,{\cal X}_2\}=-2{\cal K}^3,
\end{gather*}
and there is the functional relation $ 4{\cal H}{\cal X}_2+{\cal X}_1^2+{\cal K}^4=0$.
The degenerate potential is $V(x,y)=\frac{b_1}{x}+b_2$.

As shown in~\cite{KKW2002}, a~possible embedding of this system in 3-dimensional Euclidean space with Cartesian
coordinates $X$, $Y$, $Z$ is
\begin{gather*}
X=2\sqrt{x}\cos y,
\qquad
Y= 2\sqrt{x}\sin y,
\qquad
Z=\frac23\left(F\left(\phi,\frac{1}{\sqrt{2}}\right)+\sqrt{4x^3-x}\right),
\end{gather*}
where $x\ge\frac12$,~$y$ is $2\pi$-periodic and $\sin\phi=\sqrt{2x+1}$.
Here $F(\phi,k)$ is an incomplete elliptic integral of the 1st kind.
Then $ds^2=4x(dx^2+dy^2)=dX^2+dY^2+dZ^2$.

A general 2nd order symmetry, mod $\cal H$, can be written as ${\cal L}_1=a_1{\cal X}_2+a_2{\cal X}_1+a_3{\cal K}^2$,
and the translation group generated by $\cal K$: $x\to x$, $y\to y+\alpha$, leaves ${\cal K}^2$ and $\cal H$ invariant,
but ${\cal X}_1\to {\cal X}_1 -2\alpha {\cal H}$, ${\cal X}_2\to {\cal X}_2+\alpha {\cal X}_1-\alpha^2{\cal H}$.

We classify the distinct free nondegenerate systems under this conjugacy action.
We choose one generator ${\cal L}_1$ and determine the possibilities for ${\cal L}_2$ such that ${\cal L}_1$, ${\cal L}_2$, ${\cal H}$
form a~quadratic algebra.
Then we eliminate redundancies.
Under conjugacy we can assume that ${\cal L}_1$ takes one of the 3 forms ${\cal X}_1+a{\cal K}^2$, ${\cal X}_2+a{\cal
K}^2$, ${\cal K}^2$.

{\bf 1st case.} We choose ${\cal L}_1={\cal X}_2+a{\cal K}^2$ and try to determine the possibilities for
${\cal L}_2$, up to conjugacy under $e^{\alpha {\cal K}}$, such that ${\cal L}_1$, ${\cal L}_2$, ${\cal H}$ form a~quadratic
algebra.
(As we go through the cases step by step, we ignore systems that have already been exhibited in earlier steps.) In
general ${\cal L}_2=c_1{\cal X}_1+c_3{\cal K}^2$ and $c_1$, $c_3$ are to be determined.
We must require that
\begin{gather*}
{\cal R}^2 = b_1{\cal L}_1^3+b_2{\cal L}_2^3+b_3{\cal H}^3+b_4{\cal L}_1^2{\cal L}_2+b_5{\cal L}_1{\cal L}_2^2+b_6{\cal
L}_1{\cal L}_2{\cal H}
\\
\phantom{{\cal R}^2 =}{}
 + b_7{\cal L}_1^2{\cal H}+b_8{\cal L}_2^2{\cal H}+b_9{\cal H}^2{\cal L}_1+b_{10}{\cal H}^2{\cal L}_2,
\end{gather*}
for some constants $b_1,\dots, b_{10}$.
There are 2 possible classes:
\begin{enumerate}\itemsep=0pt
\item ${\bf \tilde D1A}$:\ ${\cal L}_1={\cal X}_2+b{\cal K}^2$, ${\cal L}_2={\cal X}_1+i{\cal K}^2$, ${\cal R}^2=2i{\cal
L}_2^3+8i{\cal L}_1{\cal L}_2{\cal H}+4b{\cal L}_2^2{\cal H}+16b{\cal H}^2{\cal L}_1$.
This class is superintegrable with
\begin{gather*}
A^{12}=0,
\qquad
A^{22}=\frac{2}{x},
\qquad
B^{12}=-\frac12\frac{5x-2b+2iy}{x(x-b+iy)},
\qquad
B^{22}=\frac{-3i}{x-b+iy},
\\
G(x,y)=\ln (4x),
\qquad
D=-\frac12(x-b+iy).
\end{gather*}
The potential of the superintegrable system is
\begin{gather*}
V(x,y)=\frac{b_1(2x-2b+iy)}{x\sqrt{x-b+iy}}+\frac{b_2}{x\sqrt{x-b+iy}}+\frac{b_3}{x}+b_4.
\end{gather*}
(This is missing in the tabulation in~\cite{KKW2002}, but pointed out in~\cite{KKMW} and~\cite{Kress2007}.)

\item ${\bf \tilde D1B}$:  ${\cal L}_1={\cal X}_2$, ${\cal L}_2={\cal K}^2$, ${\cal R}^2=-4{\cal L}_2^3-16{\cal
L}_1{\cal L}_2{\cal H}$,
\begin{gather*}
A^{12}=0,
\qquad
A^{22}=\frac{2}{x},
\qquad
B^{12}=-\frac{1}{x},
\qquad
B^{22}=-\frac{3}{y},
\\
G(x,y)=\ln(4x),
\qquad
D=-\frac{y}{2}.
\end{gather*}
(Listed as a~superintegrable system in the tabulation in~\cite{KKW2002}.) The potential of the superintegrable system is
\begin{gather*}
V(x,y)=\frac{b_1\big(4x^2+y^2\big)}{x}+\frac{b_2}{x}+\frac{b_3}{xy^2}+b_4.
\end{gather*}

{\bf 2nd case.} We choose ${\cal L}_1={\cal X}_1$.
Then the only possibility is
\item ${\bf \tilde D1C}$: ${\cal L}_1={\cal X}_1$, ${\cal L}_2={\cal K}^2$, ${\cal R}^2=16{\cal L}_2{\cal H}^2$,
\begin{gather*}
A^{12}=0,
\qquad
A^{22}=\frac{2}{x},
\qquad
B^{12}=-\frac{1}{x},
\qquad
B^{22}=0,
\\
G(x,y)=\ln(4x),
\qquad
D=-\frac{1}{2}.
\end{gather*}
(Listed in the tabulation in~\cite{KKW2002}.) The potential of the superintegrable system is
\begin{gather*}
V(x,y)=\frac{b_1\big(x^2+y^2\big)}{x}+\frac{b_2}{x}+\frac{b_3y}{x}+b_4.
\end{gather*}
\end{enumerate}

\subsubsection{Free Darboux 2 systems}
\label{Darboux2a}

The space Darboux 2 ($D_2$) has free degenerate Hamiltonian
\begin{gather*}
{\bf \tilde D2D}\colon \quad  {\cal H}=\frac{x^2}{x^2+1}\big(p_x^2+p_y^2\big)
\end{gather*}
with a~single Killing vector ${\cal K}=p_y$ and a~basis, $\{{\cal H}, {\cal K}^2, {\cal X}_1, {\cal X}_2\}$ for the
4-dimensional space of 2nd order Killing tensors.
Here,
\begin{gather*}
{\cal X}_1=2xp_xp_y+\frac{2y}{x^2+1}\big(p_y^2-x^2p_x^2\big),
\qquad
{\cal X}_2=2xyp_xp_y+\frac{\big(y^2-x^4\big)p_y^2+x^2\big(1-y^2\big)p_x^2}{x^2+1}.
\end{gather*}
The commutation relations are
\begin{gather*}
\{{\cal K},{\cal X}_1\}=2\big({\cal K}^2-{\cal H}\big),
\qquad
\{{\cal K},{\cal X}_2\}={\cal X}_1,
\qquad
\{{\cal X}_1,{\cal X}_2\}=4{\cal K}{\cal X}_2,
\end{gather*}
and there is the functional relation $ 4{\cal H}{\cal X}_2+{\cal X}_1^2-4{\cal K}^2{\cal X}_2-4{\cal H}^2=0$.
The degenerate potential is
\begin{gather*}
V(x,y)=\frac{b_1}{x^2+1}+b_2.
\end{gather*}

As shown in~\cite{KKMW} The line element $ds^2$ can be realized as a~two-dimensional surface embedded in three
dimensions~by
\begin{gather*}
X = \frac{y\sqrt{x^2+1}}x,
\qquad
Y-Z = \frac{\sqrt{x^2+1}}x,
\\
Y+Z = -\frac{\big(2x^4+5x^2+8y^2\big)\sqrt{x^2+1}}{8x} - \frac38\operatorname{arcsinh} x,
\end{gather*}
in which case,
\begin{gather*}
ds^2 = dX^2+dY^2-dZ^2 = \frac{x^2+1}{x^2}\big(dx^2+dy^2\big).
\end{gather*}

A general 2nd order symmetry, mod $\cal H$, can be written as ${\cal L}_1=a_1{\cal X}_2+a_2{\cal X}_1+a_3{\cal K}^2$ and
the translation group generated by $\cal K$: $x\to x$, $y\to y+\alpha$, leaves ${\cal K}^2$ and $\cal H$ invariant, but
\begin{gather*}
{\cal X}_1\to {\cal X}_1 -2\alpha {\cal H}+2\alpha {\cal K}^2,
\qquad
{\cal X}_2\to {\cal X}_2+\alpha {\cal X}_1+\alpha^2\big({\cal K}^2-{\cal H}\big).
\end{gather*}
We classify the distinct free nondegenerate superintegrable systems under this conjugacy action.
We choose one generator ${\cal L}_1$ and determine the possibilities for ${\cal L}_2$ such that ${\cal L}_1$, ${\cal L}_2$, ${\cal H}$
form a~quadratic algebra.
Under conjugacy there are 3 possible choices: ${\cal L}_1={\cal X}_2+a{\cal K}^2$, ${\cal X}_1$, ${\cal K}^2$.

{\bf 1st case.} We choose ${\cal L}_1={\cal X}_2+a{\cal K}^2$ and try to determine the possibilities for
${\cal L}_2$, up to conjugacy under $e^{\alpha{\cal K}}$ such that ${\cal L}_1$, ${\cal L}_2$, ${\cal H}$ form a~quadratic
algebra.
(As we go through the cases step by step, we ignore systems that have already been exhibited in earlier steps.) In
general ${\cal L}_2=c_1{\cal X}_1+c_3{\cal K}^2$ and $c_1$, $c_3$ are to be determined.
We must require that
\begin{gather*}
{\cal R}^2 = b_1{\cal L}_1^3+b_2{\cal L}_2^3+b_3{\cal H}^3+b_4{\cal L}_1^2{\cal L}_2+b_5{\cal L}_1{\cal L}_2^2+b_6{\cal L}_1{\cal L}_2{\cal H}
\\
\phantom{{\cal R}^2 =}{}
 + b_7{\cal L}_1^2{\cal H}+b_8{\cal L}_2^2{\cal H}+b_9{\cal H}^2{\cal L}_1+b_{10}{\cal H}^2{\cal L}_2,
\end{gather*}
for some constants $b_1,\dots, b_{10}$.
There are 2 possible classes:
\begin{enumerate}\itemsep=0pt
\item ${\bf \tilde D2C}$:
${\cal L}_1={\cal X}_2$, ${\cal L}_2={\cal X}_1$, ${\cal R}^2=4{\cal L}_1{\cal
L}_2^2+16{\cal L}_1^2{\cal H}-16{\cal L}_1{\cal H}^2$,
\begin{gather*}
A^{12}=0,
\qquad
A^{22}=\frac{3x^2-1}{x\big(x^2+1\big)},
\qquad
B^{12}=-\frac{\big(3x^4+x^2-2y^2\big)}{x\big(x^2+1\big)\big(x^2+y^2\big)},
\\
B^{22}=-\frac{6y}{x^2+y^2}, \qquad  G(x,y)=\ln\left(\frac{x^2+1}{x^2}\right),
\qquad
D=-x\big(x^2+y^2\big).
\end{gather*}
The potential of the superintegrable system is
\begin{gather*}
V(x,y)=\frac{x^2}{\sqrt{x^2+y^2}(x^2+1)}\left(b_1+\frac{b_2}{y+\sqrt{x^2+y^2}}+\frac{b_3}{y-\sqrt{x^2+y^2}}\right)+b_4.
\end{gather*}
(Listed as a~superintegrable system in~\cite{KKMW}.)

\item ${\bf \tilde D2B}$:  ${\cal L}_1={\cal X}_2$, ${\cal L}_2={\cal K}^2$, ${\cal R}^2=16{\cal L}_1{\cal
L}_2^2-16{\cal L}_1{\cal L}_2{\cal H}+16{\cal L}_2{\cal H}^2$,
\begin{gather*}
A^{12}=0,
\qquad
A^{22}=\frac{3x^2-1}{x\big(x^2+1\big)},
\qquad
B^{12}=-\frac{2}{x\big(x^2+1\big)},
\qquad
B^{22}=-\frac{3}{y},
\\
G(x,y)=\ln\left(\frac{x^2+1}{x^2}\right),
\qquad
D=-xy.
\end{gather*}
The potential of the superintegrable system is
\begin{gather*}
V(x,y)=\frac{x^2}{x^2+1}\left(b_1\big(x^2+y^2\big)+\frac{b_2}{x^2}+\frac{b_3}{y^2}\right)+b_4.
\end{gather*}

(Listed as a~superintegrable system in~\cite{KKMW}.)

{\bf 2nd case.} We choose ${\cal L}_1={\cal X}_1$.
Then the only possibility is
\item ${\bf \tilde D2A}$:\ ${\cal L}_1={\cal X}_1$, ${\cal L}_2={\cal K}^2$, ${\cal R}^2=16{\cal L}_2^3-32{\cal
L}_2^2{\cal H}+16{\cal L}_2{\cal H}^2$,
\begin{gather*}
A^{12}=0,
\qquad
A^{22}=\frac{3x^2-1}{x\big(x^2+1\big)},
\qquad
B^{12}=\frac{2}{x\big(x^2+1\big)},
\qquad
B^{22}=0,
\\
G(x,y)=\ln\left(\frac{x^2+1}{x^2}\right),
\qquad
D=-x.
\end{gather*}
The potential of the superintegrable system is
\begin{gather*}
V(x,y)=\frac{x^2}{x^2+1}\left(b_1(x^2+4y^2)+\frac{b_2}{x^2}+b_3y\right)+b_4.
\end{gather*}
(Listed as a~superintegrable system in~\cite{KKMW}.)
\end{enumerate}

\subsubsection{Free Darboux 3 systems}

The space Darboux 3 ($D_3$) has free degenerate Hamiltonian
\begin{gather*}
{\bf \tilde D3E}\colon \quad  {\cal H}=\frac12\frac{e^{2x}}{e^x+1}\big(p_x^2+p_y^2\big)
\end{gather*}
with a~single Killing vector ${\cal K}=p_y$ and a~basis, $\{{\cal H}, {\cal K}^2, {\cal X}_1, {\cal X}_2\}$ for the
4-dimensional space of 2nd order Killing tensors.
Here,
\begin{gather*}
{\cal X}_1=\frac12 e^x\sin y\ p_xp_y+\frac14\frac{e^{2x}}{e^x+1}\cos y\ p_x^2-\frac14 \frac{e^x(e^x+2)}{e^x+1}\cos y\
p_y^2,
\\
{\cal X}_2=-\frac12 e^x\cos y\ p_xp_y+\frac14\frac{e^{2x}}{e^x+1}\sin y\ p_x^2-\frac14 \frac{e^x(e^x+2)}{e^x+1}\sin y\
p_y^2.
\end{gather*}
The commutation relations are
\begin{gather*}
\{{\cal K},{\cal X}_1\}=-{\cal X}_2,
\qquad
\{{\cal K},{\cal X}_2\}={\cal X}_1,
\qquad
\{{\cal X}_1,{\cal X}_2\}=\frac12{\cal
K}{\cal H},
\end{gather*}
and there is the functional relation ${\cal X}_1^2+{\cal X}_2^2-\frac14{\cal H}^2-\frac12 {\cal K}^2{\cal H}=0$.
The degenerate potential is
\begin{gather*}
V(x,y)=\frac{b_1}{e^x+1}+b_2.
\end{gather*}

As shown in~\cite{KKMW}, we can embed $D3$ as a~surface in 3D Minkowski space with coordinates $X$, $Y$, $Z$ in such a~way as
to preserve rotational symmetry.
Let
\begin{gather*}
X=2\sqrt{2}e^{-\frac{x}{2}}\sqrt{1+e^{-x}}\cos \frac{y}{2},
\qquad
Y=2\sqrt{2}e^{-\frac{x}{2}}\sqrt{1+e^{-x}}\sin \frac{y}{2},
\\
Z=\frac{\sqrt{6}}{12}\ln\left(\frac{\frac{\sqrt{3}}{6}(6+5e^x){\sqrt{3+2e^{2x}+5e^x}+1}}{\frac{\sqrt{3}}{6}(6+5e^x)\sqrt{3+2e^{2x}+5e^x}-1}
\right)-e^{-x}\sqrt{2}\sqrt{3+2e^{2x}+5e^x}.
\end{gather*}
Then
\begin{gather*}
dX^2+dY^2-dZ^2=\frac{2(e^x+1)}{e^{2x}}\big(dx^2+dy^2\big).
\end{gather*}

An alternate basis is $\{{\cal H}, {\cal K}^2, {\cal Y}_1, {\cal Y}_2\}$, where
\begin{gather*}
{\cal Y}_1=\left(e^xp_xp_y+\frac{i}{2}\frac{e^{2x}}{e^x+1}p_x^2-\frac{i}{2}\frac{e^x(e^x+2)}{e^x+1}p_y^2\right)e^{iy},
\\
{\cal Y}_2=\left(e^xp_xp_y-\frac{i}{2}\frac{e^{2x}}{e^x+1}p_x^2+\frac{i}{2}\frac{e^x(e^x+2)}{e^x+1}p_y^2\right)e^{-iy},
\end{gather*}
and ${\cal X}_1=-\frac{i}{4}({\cal Y}_1-{\cal Y}_2)$, ${\cal X}_2=-\frac14({\cal Y}_1+{\cal Y}_2)$.
The new commutation relations are
\begin{gather*}
\{{\cal K},{\cal Y}_1\}=i{\cal Y}_1,
\qquad
\{{\cal K},{\cal Y}_2\}=-i{\cal Y}_2,
\qquad
\{{\cal Y}_1,{\cal Y}_2\}=-4i {\cal K{}\cal
H},
\end{gather*}
and the functional relation is ${\cal Y}_1{\cal Y}_2-{\cal H}^2-2 {\cal K}^2{\cal H}=0$.

Returning to the f\/irst basis, we note that a~general 2nd order symmetry, mod $\cal H$, can be written as ${\cal
L}_1=a_1{\cal X}_2+a_2{\cal X}_1+a_3{\cal K}^2$.
and the translation group generated by $\cal K$: $x\to x$, $y\to y+\alpha$, leaves ${\cal K}^2$ and $\cal H$ invariant,
but ${\cal X}_1\to \cos \alpha {\cal X}_1 -\sin\alpha {\cal X}_2$, ${\cal X}_2\to \sin\alpha {\cal X}_1+\cos \alpha
{\cal X}_2$.
We classify the distinct free superintegrable systems under this conjugacy action.
We choose one generator ${\cal L}_1$ and determine the possibilities for ${\cal L}_2$ such that ${\cal L}_1$, ${\cal L}_2$, ${\cal H}$
form a~quadratic algebra.
Under conjugacy the choices are ${\cal L}_1={\cal X}_1+a{\cal K}^2$, ${\cal X}_1+i {\cal X}_2$, ${\cal X}_1+i{\cal
X}_2-{\cal K}^2$, ${\cal X}_1$, ${\cal K}^2$.

{\bf 1st case.} We choose ${\cal L}_1={\cal X}_1+a{\cal K}^2$ and try to determine the possibilities for
$a$, ${\cal L}_2$, up to conjugacy under $e^{\alpha {\cal K}}$ such that ${\cal L}_1$, ${\cal L}_2$, ${\cal H}$ form a~quadratic
algebra.
(As we go through the cases step by step, we ignore systems that have already been exhibited in earlier steps.)

There are 3 possible classes:

{\bf 1st class.}
\begin{enumerate}\itemsep=0pt
\item ${\bf \tilde D3A}$:\ ${\cal L}_1={\cal X}_1-\frac12 {\cal K}^2$, ${\cal L}_2={\cal X}_2+\frac{i}{2}{\cal K}^2$.
${\cal R}^2=-{\cal L}_1^3+i{\cal L}_2^3-\frac18{\cal H}^3+i{\cal L}_1^2{\cal L}_2-{\cal L}_1{\cal L}_2^2+\frac12 ({\cal
L}_1^2+{\cal L}_2^2){\cal H} +\frac{1}{2}{\cal L}_1{\cal H}^2-\frac{i}{4}{\cal L}_2{\cal H}^2$.
Here,
\begin{gather*}
A^{12}=0,
\qquad
A^{22}=-\frac{e^x(-e^x+e^{-2x-iy}-e^{-iy}+e^{-x})}{(1+e^{-x})(e^x+1)(e^x+e^{-iy})},
\qquad
B^{22}=\frac{3ie^{-iy}}{e^x+e^{-iy}},
\\
B^{12}=\frac12\frac{e^x(-1+2e^{-x-iy}+e^{-x}+4e^{-2x-iy})} {(1+e^{-x})(e^x+e^{-iy})},
\\
G(x,y)=\ln\big(2\big(e^{-x}+e^{-2x}\big)\big),
\qquad
D=\frac{e^x}{8}\big(e^x-2e^{iy}\big).
\end{gather*}
The potential of the superintegrable system is
\begin{gather*}
V(x,y)=\frac{b_1}{1+e^x}+\frac{b_2e^x}{\sqrt{1+2e^{x+iy}}(1+e^x)}+\frac{b_3e^{x+iy}}{\sqrt{1+2e^{x+iy}}(1+e^x)}+b_4.
\end{gather*}
(This is a~superintegrable system listed in~\cite{KKMW}.)
\item ${\bf \tilde D3B}$: ${\cal L}_1={\cal X}_1$, ${\cal L}_2={\cal X}_2$, ${\cal R}^2=-\frac18 {\cal
H}^3+\frac12({\cal L}_1^2+{\cal L}_2^2){\cal H}$,
\begin{gather*}
A^{12}=0,
\qquad
A^{22}=\frac{1-e^{-x}}{1+e^{-x}},
\qquad
B^{22}=0,
\qquad
B^{12}=-\frac12\frac{(1-e^{-x})} {(1+e^{-x})},
\\
G(x,y)=\ln\big(2\big(e^{-x}+e^{-2x}\big)\big),
\qquad
D=\frac{e^{2x}}{8}.
\end{gather*}
The potential of the superintegrable system is
\begin{gather*}
V(x,y)=\frac{e^x}{e^x+1}\left(b_1+e^{-\frac{x}{2}}\left(b_2\cos\frac{y}{2}+b_3\sin\frac{y}{2}\right)\right)+b_4.
\end{gather*}
(Listed as a~superintegrable system in~\cite{KKMW}.)
\item ${\bf \tilde D3C}$:\ ${\cal L}_1={\cal X}_1$, ${\cal L}_2={\cal K}^2$, ${\cal R}^2=-4{\cal L}_1^2{\cal L}_2+2{\cal
L}_2^2{\cal H}+{\cal L}_2{\cal H}^2$,
\begin{gather*}
A^{12}=0,
\qquad
A^{22}=\frac{1-e^{-x}}{1+e^{-x}},
\qquad
B^{12}=\frac{1+2e^{-x}}{1+e^{-x}},
\qquad
B^{22}=-3\cot y,
\\
G(x,y)=\ln(2(e^{-x}+e^{-2x})),
\qquad
D=-\frac{e^{x}}{4}\sin y.
\end{gather*}
The potential of the superintegrable system is
\begin{gather*}
V(x,y)= \frac{e^x}{e^x+1}\left(b_1+e^x\left(\frac{b_2}{\cos^2\frac{y}{2}}+\frac{b_3}{\sin^2\frac{y}{2}}\right)\right)+b_4.
\end{gather*}
(Listed as a~superintegrable system in~\cite{KKMW}.)
\end{enumerate}

{\bf 2nd class.} We choose ${\cal L}_1={\cal X}_1+i{\cal X}_2$.
Then the only new possibility is
\begin{enumerate}\itemsep=0pt
\setcounter{enumi}{3}
\item ${\bf \tilde D3D}$:\ ${\cal L}_1={\cal X}_1+i{\cal X}_2$, ${\cal L}_2={\cal K}^2$, ${\cal R}^2=-4{\cal L}_1^2{\cal
L}_2$,
\begin{gather*}
A^{12}=0,
\qquad
A^{22}=\frac{1-e^{-x}}{1+e^x},
\qquad
B^{12}=\frac{1+2e^{-x}}{1+e^{-x}},
\qquad
B^{22}=-3i,
\\
G(x,y)=\ln(2\big(e^{-x}+e^{-2x}\big)),
\qquad
D=\frac{i}{4}e^{x+iy}.
\end{gather*}
The potential of the superintegrable system is
\begin{gather*}
V(x,y)=\frac{e^{2x}}{1+e^x}\big(b_1 e^{-iy}+b_2 e^{-2iy}\big)+\frac{b_3}{1+e^x}+b_4.
\end{gather*}
(Listed as a~superintegrable system in~\cite{KKMW}.)
\end{enumerate}

{\bf 3rd class.} We choose ${\cal L}_1={\cal X}_1+i{\cal X}_2-{\cal K}^2$.
There are no new possibilities.

\subsubsection{Free Darboux 4 systems}

The spaces Darboux 4 ($D_4(b)$) have free degenerate Hamiltonian
\begin{gather}
\label{Darboux4}
{\bf \tilde D4(b)D}\colon \quad  {\cal H}=-\frac{\sin^2 2x}{2\cos 2x+b}\big(p_x^2+p_y^2\big),
\end{gather}
$b\ne\pm 2$, with a~single Killing vector ${\cal K}=p_y$ and a~basis, $\{{\cal H},{\cal K}^2,{\cal X}_1,{\cal X}_2\}$
for the 4-dimensional space of 2nd order Killing tensors.
Here,
\begin{gather*}
{\cal X}_1= e^{2y}\big({-}{\cal H}+\cos 2x p_y^2+\sin 2x p_xp_y\big),
\qquad
{\cal X}_2= e^{-2y}\big({-}{\cal H}+\cos 2x p_y^2-\sin 2x p_xp_y\big).
\end{gather*}
The commutation relations are
\begin{gather*}
\{{\cal K},{\cal X}_1\}=2{\cal X}_1,
\qquad
\{{\cal K},{\cal X}_2\}=-2{\cal X}_2,
\qquad
\{{\cal X}_1,{\cal X}_2\}=-8{\cal
K}^3-4b{\cal K}{\cal H},
\end{gather*}
and there is the functional relation ${\cal X}_1{\cal X}_2-{\cal K}^4-b{\cal K}^2{\cal H}-{\cal H}^2=0$.
The degenerate potential is
\begin{gather*}
V(x,y)=\frac{b_1}{2\cos 2x+b}+b_2.
\end{gather*}

A general 2nd order symmetry, mod $\cal H$, can be written as ${\cal L}_1=a_1{\cal X}_2+a_2{\cal X}_1+a_3{\cal K}^2$,
and the translation group generated by $\cal K$: $x\to x$, $y\to y+\alpha$, leaves ${\cal K}^2$ and $\cal H$ invariant,
but ${\cal X}_1\to e^{-2\alpha} {\cal X}_1$, ${\cal X}_2\to e^{2\alpha} {\cal X}_2$.
Also the ref\/lection $y\to -y$, $x\to x$ leaves $\cal H$ and ${\cal K}^2$ f\/ixed but ${\cal X}_1\leftrightarrow {\cal
X}_2$.
We classify the distinct free superintegrable systems under this conjugacy action.
We choose one generator ${\cal L}_1$ and determine the possibilities for ${\cal L}_2$ such that ${\cal L}_1$, ${\cal L}_2$, ${\cal H}$
form a~quadratic algebra.
Under conjugacy the choices are ${\cal L}_1={\cal K}^2+a{\cal X}_2$, ${\cal X}_1+{\cal X}_2+a{\cal K}^2$ ($a=0,2$,
or $a\ne 0,2$), ${\cal X}_2$, ${\cal K}^2$.

{\bf 1st case.} We choose ${\cal L}_1={\cal K}^2+a{\cal X}_2$ and try to determine the possibilities for
${\cal L}_2$, up to conjugacy under $e^{\alpha{\cal K}}$ such that ${\cal L}_1$, ${\cal L}_2$, ${\cal H}$ form a~quadratic
algebra.
(As we go through the cases step by step, we ignore systems that have already been exhibited in earlier steps.) We f\/irst
try ${\cal L}_2={\cal X}_2+c{\cal X}_1$.
The class $ac\ne 0$ doesn't yield a~quadratic algebra.
There are 3 other possible classes:
\begin{enumerate}\itemsep=0pt
\item ${\bf \tilde D4(b)A}$: ${\cal L}_1={\cal K}^2$, ${\cal L}_2={\cal X}_2+{\cal X}_1$, ${\cal R}^2=-64{\cal
L}_1^3+16{\cal L}_1{\cal L}_2^2-64b{\cal L}_1^2{\cal H}-64{\cal L}_1{\cal H}^2$,
\begin{gather*}
A^{12}=0,
\qquad
A^{22}=\frac{2(-2\sin^2 2x-b\cos 2x-2)}{\sin 2x(2\cos 2x+b)},
\\
B^{12}=\frac{4(-\sin^2 2x+b\cos 2x+2)}{\sin 2x (2\cos 2x+b)},
\qquad
B^{22}=6\left(\frac{1+e^{4y}}{1-e^{4y}}\right),
\\
G(x,y)=\ln\left(\frac{-b-2\cos 2x}{\sin^2 2x}\right),
\qquad
D=\frac{\sin 2x}{2}\big(e^{2y}+e^{-2y}\big).
\end{gather*}
The potential of the superintegrable system is
\begin{gather*}
V(x,y)=\frac{\sin^2 2 x}{2 \cos 2 x+b}\left(\frac{b_1}{\sinh^2 y}+\frac{b_2}{\sinh^2 2 y}\right)+\frac{b_3}{2 \cos 2
x+b}+b_4.
\end{gather*}
(Listed as a~superintegrable system in~\cite{KKMW}.)
\item ${\bf \tilde D4(b)B}$:\ ${\cal L}_1={\cal K}^2$, ${\cal L}_2={\cal X}_2$, ${\cal R}^2=16{\cal L}_1{\cal L}_2^2$,
\begin{gather*}
A^{12}=0,
\qquad
A^{22}=\frac{2(-2\sin^2 2x-b\cos 2x-2)}{\sin 2x(2\cos 2x+b)},
\\
B^{12}=\frac{4(-\sin^2 2x+b\cos 2x+2)}{\sin
2x (2\cos 2x+b)},
\qquad
B^{22}=6,
\\
G(x,y)=\ln\left(\frac{-b-2\cos 2x}{\sin^2 2x}\right),
\qquad
D=-\frac{\sin 2x}{2}e^{-2y}.
\end{gather*}
The potential of the superintegrable system is
\begin{gather*}
V(x,y)=\frac{\sin^2 2x}{2 \cos 2x +b}\left(\frac{b_1}{\sin^2 2x}+b_2e^{4y}+b_3e^{2y}\right)+b_4.
\end{gather*}
(Listed as a~superintegrable system in~\cite{KKMW}.)
\end{enumerate}

{\bf 2nd case.} We choose ${\cal L}_1={\cal K}^2+a{\cal X}_2$, ${\cal L}_2={\cal X}_1$.
Then the only possibility is $a=0$, which is redundant.

{\bf 3rd case.} We choose ${\cal L}_1={\cal X}_1+{\cal X}_2+a{\cal K}^2$, ${\cal L}_2={\cal X}_2+c{\cal K}^2$.
We generate a~quadratic algebra for the system
\begin{enumerate}\itemsep=0pt
\setcounter{enumi}{2}
\item ${\bf \tilde D4(b)C}$:\ ${\cal L}_1={\cal X}_1+{\cal X}_2+2{\cal K}^2$, ${\cal L}_2={\cal X}_2+{\cal K}^2$.
${\cal R}^2= -16b{\cal H}^3+16{\cal L}_1^2{\cal L}_2-16{\cal L}_1{\cal L}_2^2+16b{\cal L}_1{\cal L}_2{\cal H}-16b{\cal
L}_2^2{\cal H}-16{\cal L}_1{\cal H}^2$,
\begin{gather*}
A^{12}=0,
\qquad
A^{22}=-\frac{4\sin^2 2x +2b\cos 2x+4}{\sin 2x(2\cos 2x +b)},
\qquad
B^{22}=-\frac{6(e^{4y}-1)}{2e^{2y}\cos
2x+e^{4y}+1},
\\
B^{12}=\frac{2be^{2y}\cos^2 2x +6b e^{2y}+4b\cos 2x +4be^{4y}\cos 2x -8\cos 2x +16\cos^4 x} {\sin 2x (e^{4y}+2e^{2y}\cos 2x+1)(2\cos 2x +b)},
\\
\phantom{B^{12}=}{}
+\frac{-8e^{4y}\cos 2x+16e^{4y}\cos^4 x-4e^{2y}\cos^3 2x+16 e^{2y}\cos 2x}{\sin 2x (e^{4y}+2e^{2y}\cos 2x+1)(2\cos 2x +b)},
\\
G(x,y)=\ln\left(\frac{-b-2\cos 2x}{\sin^2 2x}\right),
\qquad
D=-\frac{\sin 2x}{2}\big(2\cos 2x+e^{-2y}+e^{2y}\big).
\end{gather*}
The potential of the superintegrable system is
\begin{gather*}
V(x,y)=\frac{e^{2y}}{\frac{b+2}{\sin^2 x}+\frac{b-2}{\cos^2
x}}\left(\frac{b_1}{Z+(1-e^{2y})\sqrt{Z}}+\frac{b_2}{Z+(1+e^{2y})\sqrt{Z}} +\frac{b_3\ e^{-2y}}{\cos^2 x}\right) +b_4,
\end{gather*}
where $Z=(1-e^{2y})^2+4e^{2y}\cos^2 x$.
(Listed as a~superintegrable system in~\cite{KKMW}.)
\end{enumerate}

{\bf 4th case.} We choose ${\cal L}_1={\cal X}_1$, ${\cal L}_2={\cal X}_2$.
We do not generate a~quadratic algebra.

\subsection{Alternate free Darboux 4 systems}

There is an alternate form of $\tilde D4(b)$ that we shall employ.
Set
\begin{gather*}
x=iX,
\qquad
y=iY,
\qquad
{\cal J}=i{\cal K}=p_Y,
\qquad
{\cal Y}_1=\frac{{\cal X}_1+{\cal X}_2}{2},
\qquad
{\cal Y}_2=\frac{{\cal X}_1-{\cal X}_2}{2i}.
\end{gather*}
Then for $b\ne\pm 2$ we have
\begin{gather}
{\cal H}=\frac{\sinh^2 2X}{2\cosh 2X+b}\big(p_X^2+p_Y^2\big),
\nonumber
\\
{\cal Y}_1=-\cos(2Y)\big({\cal H}-\cosh 2X \, p_Y^2\big)-\sin (2Y)\sinh 2X\, p_Xp_Y,
\nonumber
\\
{\cal Y}_2=-\sin(2Y)\big({\cal H}-\cosh 2X \, p_Y^2\big)+\cos (2Y)\sinh 2X\, p_Xp_Y.
\label{Darboux4b}
\end{gather}
The structure equations are
\begin{gather*}
\{{\cal J},{\cal Y}_1\}=-2{\cal Y}_2,
\qquad
\{{\cal J},{\cal Y}_2\}=2{\cal Y}_1,
\qquad
\{{\cal Y}_1,{\cal Y}_2\}=4{\cal J}^3+2b{\cal J}{\cal H},
\\
{\cal Y}_1^2+{\cal Y}_2^2-{\cal J}^4-b{\cal J}^2{\cal H}-{\cal H}^2=0.
\end{gather*}
We can embed $D4(b)$ as a~surface in 3D Minkowski space with Cartesian coordinates~$X$,~$Y$,~$Z$.
Let
\begin{gather*}
X=y\sqrt{b+2\cosh 2x},
\qquad
Y-Z=\sqrt{b+2\cosh 2x},
\\
Y+Z=\frac{y^2\sqrt{b+2\cosh 2x}}{\sinh 2x}+\int\frac{2\cosh 4x+2+4b\cosh 2x +b^2}{\sqrt{b+2\cosh 2x}(\cosh
4x+3+2b\cosh 2x)} dx.
\end{gather*}

Then
\begin{gather*}
ds^2=\frac{2\cosh 2x+b}{\sinh^2 2x}\big(dx^2+dy^2\big)=-dX^2+dY^2-dZ^2.
\end{gather*}
The change of variable $u=e^x$ converts the integral into an elliptic integral in~$u$ that can be evaluated as a~rather
complicated sum of elementary functions and the elliptic integrals of types one, two and three.

In terms of this alternate form the superintegrable systems can be expressed as:
\begin{enumerate}\itemsep=0pt
\item ${\bf \tilde D4(b)A'}$:  ${\cal L}_1={\cal J}^2$, ${\cal L}_2={\cal Y}_1$, ${\cal R}^2=16{\cal L}_1^3-16{\cal
L}_1{\cal L}_2^2+16b{\cal L}_1^2{\cal H}+16{\cal L}_1{\cal H}^2$,
\begin{gather*}
A^{12}=0,
\qquad
A^{22}=-\frac{2(-2\sinh^2 2X+b\cosh 2X+2)}{\sinh 2X(2\cosh 2X+b)},
\qquad
B^{22}=-6\left(\frac{\cos 2Y}{\sin 2Y}\right),
\\
B^{12}=\frac{4(\sinh^2 2X+b\cosh 2X+2)}{\sinh 2X (2\cosh 2X+b)}, \qquad  G(x,y)=\ln\left(\frac{b+2\cosh 2X}{\sinh^2 2X}\right).
\end{gather*}

\item ${\bf \tilde D4(b)B'}$:  ${\cal L}_1={\cal J}^2$, ${\cal L}_2={\cal Y}_1-i{\cal Y}_2$, ${\cal R}^2=-16{\cal
L}_1{\cal L}_2^2$,
\begin{gather*}
A^{12}=0,
\qquad
A^{22}=\frac{2(-2\sinh^2 2X+b\cosh 2X+2)}{\sinh 2X(2\cosh 2X+b)}, \qquad  B^{22}=6i,
\\
B^{12}=\frac{4(+\sinh^2 2X+b\cosh 2X+2)}{\sinh 2X (2\cosh 2X+b)}, \qquad  G(x,y)=\ln\left(\frac{b+2\cosh 2X}{\sinh^2 2X}\right).
\end{gather*}

\item ${\bf \tilde D4(b)C'}$:  ${\cal L}_1={\cal Y}_1-{\cal J}^2$, ${\cal L}_2={\cal Y}_2$.
${\cal R}^2= -4b{\cal H}^3-8{\cal L}_1^3-8{\cal L}_1{\cal L}_2^2+4b({\cal L}_1^2+{\cal L}_2^2){\cal H}+8{\cal L}_1{\cal
H}^2$.
\begin{gather*}
A^{12}=0,
\qquad
A^{22}=-\frac{-4\sinh^2 2X +2b\cosh 2X+4}{\sinh 2X(2\cosh 2X +b)},
\qquad
B^{22}=\frac{6\sin 2Y}{-\cosh 2X+\cos 2Y},
\\
B^{12} = -\frac{-2b\cosh^2 2X -6b +(8b+16)\cosh 2X\cos 2Y} {\sinh 2X (2\cos 2Y -2\cosh 2X)(2\cosh 2X +b)}
\\
\phantom{B^{12} =}{}
 - \frac{32\cosh^4 2X\cos 2Y+4\cosh^3 2X-16\cosh 2X} {\sinh 2X (2\cos 2Y -2\cosh 2X)(2\cosh 2X +b)},
\\
G(x,y)=\ln\left(\frac{b+2\cosh 2X}{\sinh^2 2X}\right).
\end{gather*}

\item ${\bf \tilde D4(b)D'}$
(free degenerate): ${\cal J}$, ${\cal L}_1={\cal Y}_1$, ${\cal L}_2={\cal Y}_2$, ${\cal
Y}_1^2+{\cal Y}_2^2-{\cal J}^4-b{\cal J}^2{\cal H}-{\cal H}^2=0$.
\end{enumerate}

\subsection{Summary and an example}
\label{example1}

From the results of~\cite{KM2014} and the calculations of the preceding sections we see that Theorems~\ref{theorem2}
and~\ref{theorem3} are valid for Darboux spaces.

We use $\bf D2C$ as an example to give more details about how a~nondegenerate superintegrable system with potential is
induced from a~free system: From Section~\ref{Darboux2a} we have the free Darboux 2 system
\begin{gather*}
{\cal L}_1={\cal X}_2,
\qquad
{\cal L}_2={\cal X}_1,
\qquad
{\cal R}^2=4{\cal L}_1{\cal L}_2^2+16{\cal L}_1^2{\cal H}-16{\cal L}_1{\cal H}^2.
\end{gather*}
The potential equations are determined~by
\begin{gather}
A^{12}=0,
\qquad
A^{22}=\frac{3x^2-1}{x(x^2+1)},
\nonumber
\\
B^{12}=-\frac{\big(3x^4+x^2-2y^2\big)}{x\big(x^2+1\big)\big(x^2+y^2\big)},
\qquad
B^{22}=-\frac{6y}{x^2+y^2},
\label{D3Ccan}
\end{gather}
and the general solution is
\begin{gather}
\label{D3Cpot}
V(x,y)=\frac{x^2}{2\sqrt{x^2+y^2}\big(x^2+1\big)}\left(b_1+\frac{b_2}{y+\sqrt{x^2+y^2}}+\frac{b_3}{-y+\sqrt{x^2+y^2}}\right)+b_4.
\end{gather}
The induced classical system has a~basis of symmetries
\begin{gather}
{\cal H}=\frac{x^2}{x^2+1}\big(p_x^2+p_y^2\big)+V(x,y),
\qquad
{\cal L}_1=2xyp_yp_x+\frac{\big(y^2-x^4\big)p_y^2+x^2\big(1-y^2\big)p_x^2}{x^2+1}+W^{(1)},
\nonumber
\\
\label{D3Csym}
{\cal L}_2=2xp_xp_y+\frac{2y\big(p_y^2-x^2p_x^2\big)}{x^2+1}+W^{(2)},
\end{gather}
where
\begin{gather*}
W^{(1)}=\frac12\frac{b_1y\big(1-x^2\big)+b_2\big(\big({-}y+\sqrt{x^2+y^2}\big)^2+1\big)-b_3\big(\big(y+\sqrt{x^2+y^2}\big)^2+1\big)}{\big(x^2+1\big)\sqrt{x^2+y^2}},
\\
W^{(2)}=\frac{2b_1\sqrt{x^2+y^2}-b_2\big(\big({-}y+\sqrt{x^2+y^2}\big)^2-1\big)-b_3\big(\big(y+\sqrt{x^2+y^2}\big)^2-1\big)}{4x^2+4}.
\end{gather*}
The Casimir is
\begin{gather}
{\cal R}^2 - 4{\cal L}_1{\cal L}_2^2-16{\cal L}_1^2{\cal H}+16{\cal L}_1{\cal H}^2-4(b_2+b_3)H^2+16b_4{\cal
L}_1^2+(b_2+b_3){\cal L}_2^2 -32b_4{\cal H}{\cal L}_1
\nonumber
\\
\qquad{}
 + \big(8b_2b_4-cb_1^2+4cb_2b_3+8b_3b_4\big){\cal H}+\big(16b_4^2+b_1^2\big){\cal L}_1+b_1(-b_3+b_2){\cal L}_2
\nonumber
\\
\qquad{}
 + \left(\frac14 b_1^2b_2+\frac14 b_1^2b_3+b_1^2b_4-4b_3b_4^2-4b_2b_3b_4-4b_2b_4^2\right).
\label{D3Ccas}
\end{gather}

We will not work out the details of the quantum case but merely note that the potential terms of the symmetries remain
unchanged as do the 2nd order kinetic energy terms, but there are now 1st order terms: the Hamiltonian kinetic energy is
replaced by the Laplace--Beltrami operator on D2 and the other generating symmetry operators are formally self-adjoint
with respect to the D2 volume measure $x^2\ dx \ dy /(x^2+1)$.

\section{Contractions of Darboux systems}

Recall that the scalar curvature of a~space with metric $ds^2=e^{G(x,y)}(dx^2+dy^2)$, where $\lambda(x,y)=e^{G(x,y)}$ is
$C=-e^{-G}(\partial_{xx}G+\partial_{yy}G)$.
Constant curvature spaces are just those for which~$C$ is constant; f\/lat spaces are those for which $C=0$.
\begin{theorem}
\label{Ricci1}
A~Darboux or Koenigs superintegrable system cannot be obtained as a~geometric contraction of a~nonzero constant
curvature or flat space superintegrable system.
\end{theorem}

\begin{proof}  Suppose we have a~contraction of a~nonzero constant curvature system with Hamiltonian ${\cal
H}=\frac{p_x^2+p_y^2}{e^G}$.
Then there is a~1-parameter family of Hamiltonians ${\cal H}' = \frac{p_{x'}^2+p_{y'}^2}{e^{G'(\epsilon)}}$, where
\mbox{$G'(1)=G$}, $G'$ depends smoothly on~$\epsilon$ in the interval $0\le \epsilon\le 1$, and $G'(0)$ def\/ines the metric of
the target manifold for the contraction.
Further, for $\epsilon\ne 0$ the metric def\/ined by $G'$ will be a~scalar multiple of a~metric on the original constant
curvature system.
Thus we have $ C(\epsilon)=-e^{-G'}(\partial_{x'x'}G'+\partial_{y'y'}G')$ for $\epsilon\ne 0$, where $C(\epsilon)$ is
nonzero and independent of $x'$, $y'$.
In the limit we obtain the constant $C(0)$, so the target manifold must be f\/lat or of nonzero constant curvature, hence
not a~Darboux or Koenigs manifold.
Similarly, if the original manifold is f\/lat the target manifold must also be f\/lat.
\end{proof}

Now we investigate contractions involving Darboux superintegrable systems.
From Theorem~\ref{Ricci1} such systems cannot be obtained as contractions of constant curvature systems.
(However, they are all St\"ackel equivalent to constant curvature space systems.) Thus we limit ourselves to the search
for contractions such that the originating manifold is a~Darboux space.
In distinction to the case of constant curvature spaces as originating manifolds (where all quadratic algebra
contractions were induced by Lie algebra contractions of $e(2,{\mathbb C})$ and $o(3,{\mathbb C})$) here the Darboux
spaces have only a~one-dimensional Lie symmetry algebra, so we have to search for new contraction mechanisms.
However, we can restrict our search to free systems and then induce the superintegrable system contractions
automatically.

To induce the contractions through Lie algebra methods we will employ the conformal symmetry algebra for each free
Darboux system generated by functions ${\cal Q}(x,y,p_x,p_y)=A(x,y)p_x+B(x,y)p_y$ that satisfy the relations
\begin{gather*}
\{{\cal H},
{\cal Q}\}=R_Q(x,y){\cal H},
\qquad
\{{\cal K},
{\cal Q}\}=K_Q(x,y){\cal K},
\qquad
{\cal K}=p_y.
\end{gather*}
A~straightforward calculation gives the same algebra in each case:
\begin{gather*}
{\bf{\cal G}3}\colon \  {\cal P}_x=p_x,
\qquad
{\cal P}_y=p_y,
\qquad
{\cal M}=x p_x+y p_y,
\\
\{{\cal P}_x,{\cal P}_y\}=0,
\qquad
\{{\cal P}_x,{\cal M}\}={\cal P}_x,
\qquad
\{{\cal P}_y,{\cal M}\}={\cal P}_y.
\end{gather*}
The In\"on\"u--Wigner contractions and their geometric implementations are:
\begin{enumerate}\itemsep=0pt
\item[1)] ${\cal P}_x$, ${\cal P}_y$, $\epsilon{\cal M}$;\qquad
$x=x'+\frac{1}{\epsilon}$, $y=y'$,  or $x=x'$, $y=y'+\frac{1}{\epsilon}$,
\item[2)] $\epsilon {\cal P}_x$, $\epsilon{\cal P}_y$, ${\cal M}$;
\qquad
$x=\epsilon x'$, $y=\epsilon y'$.
\end{enumerate}
Up to an isomorphism of the Lie algebra that is obtained by contraction of {\cal G}3 there are no continuous one
parametric contractions besides these two,~\cite{NP}.
Each of these contractions does induce a~corresponding contraction of each free Darboux system and we have found no
contractions of Darboux systems other than these.
This approach is compatible with the use of generalizations of In\"on\"u--Wigner contractions for the symmetry algebras
$o(3,{\mathbb C})$ and $e(2,{\mathbb C})$ of constant curvature spaces.
Indeed if we compute the conformal symmetry algebra for each free constant curvature system generated by functions
${\cal Q}(x,y,p_x,p_y)=A(x,y)p_x+B(x,y)p_y$ that satisfy the relations
\begin{gather*}
\{{\cal H}, {\cal Q}\}=R_Q(x,y){\cal H},
\qquad
\{{\cal L}_j, {\cal Q}\}=\sum\limits_{\ell=1}^3K_{Q,j}^{(\ell)}(x,y){\cal L}_\ell,
\end{gather*}
where the ${\cal L}_j$ form a~basis for either $o(3,{\mathbb C})$ or $e(2,{\mathbb C})$ we f\/ind $o(3,{\mathbb C})$ in
the case of the sphere and the af\/f\/ine algebra, the semidirect product of $e(2,{\mathbb C})$ and the dilation ${\cal M}$,
for f\/lat space.
One can show that the geometrical implementations of generalized In\"on\"u--Wigner contractions of the af\/f\/ine algebra
either agree with those of $e(2,{\mathbb C})$ or the contractions cannot be implemented geometrically.
See for example~\cite[Table~XXIX]{PH} for the ordinary In\"on\"u--Wigner contractions of the af\/f\/ine algebra.

\subsection{D1 contractions}

We list approaches to f\/inding the contractions (not the contractions themselves).
\begin{enumerate}\itemsep=0pt
\item Let $x=x'+\frac{1}{\epsilon}$, $y=y'$, ${\cal H}'=\frac{4}{\epsilon} {\cal H}$.
Then as $\epsilon \to 0$ we have
\begin{gather*}
{\cal H}'=p_{x'}^2+p_{y'}^2,
\qquad
{\cal X}_1\to p_{x'}p_{y'},
\qquad
{\cal X}_2\approx p_{y'}(y'p_{x'}-x'p_{y'})-\frac{p_{y'}^2}{\epsilon}.
\end{gather*}
Contractions constructed from such limits would have f\/lat space as the target manifold.
\item Let $x=x'$, $y=y'+\frac{1}{\epsilon}$, ${\cal H}'= {\cal H}$.
Then as $\epsilon \to 0$ we have
\begin{gather*}
{\cal H}'=\frac{p_{x'}^2+p_{y'}^2}{4x'},
\qquad
{\cal X}_1\approx p_{x'}p_{y'}-\frac{y'}{2x'}\big(p_{x'}^2+p_{y'}^2\big)-\frac{2}{\epsilon}{\cal H}',
\\
{\cal X}_2\approx p_{y'}(y'p_{x'}-x'p_{y'})+\frac{p_{x'}p_{y'}}{\epsilon}-\frac{{y'}^2}{4x'}\big(p_{x'}^2+p_{y'}^2\big)-\left(\frac{y'}{2\epsilon
x'}+\frac{1}{4\epsilon^2 x'}\right) \big(p_{x'}^2+p_{y'}^2\big).
\end{gather*}

Contractions constructed from such limits would have D1 itself as the target manifold.
\item Let $x=\epsilon x'$, $y=\epsilon y'$, ${\cal H}'=4\epsilon {\cal H}$.
Then as $\epsilon \to 0$ we have ${\cal H}'= \frac{p_{x'}^2+p_{y'}^2}{x'}$ and
\begin{gather*}
\epsilon^2 {\cal X}_1= p_{x'}p_{y'}-\frac{y'}{2x'}\big(p_{x'}^2+p_{y'}^2\big),
\\
\epsilon {\cal X}_2=p_{y'}(y'p_{x'}-x'p_{y'})-\frac{{y'}^2}{4x'}\big(p_{x'}^2+p_{y'}^2\big),
\qquad
\epsilon {\cal K}=p_{y'}.
\end{gather*}
Contractions constructed from such limits would have D1 again as the target manifold.
\end{enumerate}

\subsection{D2 contractions}

We list approaches to f\/inding the contractions (not the contractions themselves).
\begin{enumerate}\itemsep=0pt
\item Let $x=x'+\frac{1}{\epsilon}$, $y=y'$, ${\cal H}'= {\cal H}$.
Then as $\epsilon \to 0$ we have
\begin{gather*}
{\cal H}'=p_{x'}^2+p_{y'}^2,
\qquad
\epsilon {\cal X}_1\approx2 p_{x'}p_{y'},
\qquad
\epsilon^2 {\cal X}_2\approx -p_{y'}^2.
\end{gather*}
Contractions constructed from such limits would have f\/lat space as the target manifold.
\item Let $x=\epsilon x'$, $y=\epsilon y'$, ${\cal H}'={\cal H}$.
Then as $\epsilon \to 0$ we have ${\cal H}'= {x'}^2(p_{x'}^2+p_{y'}^2)$ and
\begin{gather*}
\epsilon {\cal X}_1\approx 2x'p_{x'}p_{y'}+2y'p_{y'}^2,
\qquad
\epsilon {\cal X}_2=2x'y'p_{x'}p_{y'}+{y'}^2p_{y'}^2+{x'}^2p_{x'}^2,
\qquad
\epsilon {\cal K}=p_{y'}.
\end{gather*}
Contractions constructed from such limits would have the complex 2-sphere as the target manifold.
\item Let $x= x'$, $y= y'+\frac{1}{\epsilon}$, ${\cal H}'={\cal H}$.
Then as $\epsilon \to 0$ we have ${\cal H}'= \frac{{x'}^2}{{x'}^2+1}(p_{x'}^2+p_{y'}^2)$, ${\cal K}'={\cal K}$,
\begin{gather*}
{\cal X}_1'(x',y',p_{x'},p_{y'})={\cal X}_1+\frac{2}{\epsilon}\big({\cal K}^2-{\cal H}\big),
\\
{\cal X}_2'(x',y',p_{x'},p_{y'})={\cal X}_2 +\frac{1}{\epsilon^2}\big({\cal K}^2-{\cal H}\big)+\frac{1}{\epsilon}{\cal X}_1.
\end{gather*}
Contractions constructed from such limits would have $D2$ as the target manifold.
\end{enumerate}

\subsection{D3 contractions}

We list approaches to f\/inding the contractions (not the contractions themselves).
\begin{enumerate}\itemsep=0pt
\item Let $x=x'+\ln(\frac{1}{\epsilon})$, $y=y'$, ${\cal H}'= 2\epsilon{\cal H}$.
Then as $\epsilon \to 0$ we have ${\cal K}=p_{y'}$
\begin{gather*}
{\cal H}'=e^{x'}\big(p_{x'}^2+p_{y'}^2\big),
\qquad
\epsilon {\cal X}_1\approx \frac{e^{x'}}{2}\left(\sin y'\ p_{x'}p_{y'}+\frac12\cos
y'\big(p_{x'}^2-p_{y'}^2\big)\right),
\\
\epsilon {\cal X}_2\approx \frac{e^{x'}}{2}\left(-\cos y'\ p_{x'}p_{y'}+\frac12\sin y'\big(p_{x'}^2-p_{y'}^2\big)\right).
\end{gather*}
Contractions constructed from such limits would have f\/lat space as the target manifold.
In terms of f\/lat space Cartesian coordinates $X=r\cos\theta$, $Y=r\sin\theta$ we have
\begin{gather*}
e^{x'}=\frac{4}{r^2},
\qquad
y'=2\theta,
\qquad
{\cal H}'=p_X^2+p_Y^2,
\qquad
\epsilon {\cal X}_1\approx \frac14\big(p_X^2-p_Y^2\big),
\\
\epsilon {\cal X}_2\approx \frac12p_Xp_Y,
\qquad
{\cal K}'=\frac12(Xp_Y-Yp_X)=\frac12{\cal J}.
\end{gather*}
\item Let $x=\epsilon x'$, $y=\epsilon y'$, ${\cal H}'=4\epsilon^2{\cal H}$.
Then as $\epsilon \to 0$ we have ${\cal H}'=p_{x'}^2+p_{y'}^2$ and
\begin{gather*}
\epsilon^2 {\cal X}_1\approx \frac{1}{8}\big(p_{x'}^2-3p_{y'}^2\big),
\qquad
\epsilon^2 {\cal X}_2\approx -\frac{1}{2}p_{x'}p_{y'},
\qquad
\epsilon {\cal K}=p_{y'}.
\end{gather*}
Contractions constructed from such limits would have f\/lat space as the target manifold.
\item Let $x=x'$, $y=y'+i\ln\epsilon$, ${\cal H}'={\cal H}'$.
Then as $\epsilon \to 0$ we have
\begin{gather*}
{\cal Y}_1'(x',y',p_{x'},p_{y'}) \approx \epsilon {\cal Y}_1,
\qquad
{\cal Y}_2'(x',y',p_{x'},p_{y'}) \approx \frac{1}{\epsilon} {\cal Y}_2,
\qquad
{\cal K}=p_{y'}.
\end{gather*}
Contractions constructed from such limits would have $D3$ as the target manifold.
\end{enumerate}

\subsection{D4 contractions}

We list approaches to f\/inding the contractions (not the contractions themselves).
\begin{enumerate}\itemsep=0pt
\item Let $X=X'+\ln\epsilon$, $Y=Y'$, ${\cal H}'= 4\epsilon^2{\cal H}$, with~$b$ f\/ixed.
Then as $\epsilon \to 0$ we have ${\cal J}=p_{Y'}$ and
\begin{gather*}
{\cal H}'=e^{-2X'}\big(p_{X'}^2+p_{Y'}^2\big),
\qquad
{\cal J}'=p_{Y'},
\\
4\epsilon^2 {\cal Y}_1\approx -\cos (2Y')\big({\cal H}'-2e^{-2X'}p_{Y'}^2\big)+2\sin (2Y')  e^{-2X'}p_{X'}p_{Y'}\approx
p_y^2-p_x^2,
\\
4\epsilon^2 {\cal Y}_2\approx -\sin(2Y')\big({\cal H}'-2e^{-2X'}p_{Y'}^2\big)-2\cos(2Y')e^{-2X'}p_{X'}p_{Y'}\approx -2p_xp_y,
\end{gather*}
where $x$, $y$ are standard Cartesian coordinates.
Contractions constructed from such limits would have f\/lat space as the target manifold.
\item Let $X=\epsilon X'$, $Y=\epsilon Y'$, ${\cal H}'={\cal H}$, with~$b$ f\/ixed.
Then as $\epsilon \to 0$ we have ${\cal H}'= \frac{4{X'}^2}{2+b}(p_{X'}^2+p_{Y'}^2)$ and
\begin{gather*}
{\cal Y}_1-p_Y^2+{\cal H}\approx 2\big({X'}^2-{Y'}^2\big)p_{Y'}^2-4X'Y'p_{X'}p_{Y'},
\\
\epsilon {\cal Y}_2\approx 2Y'p_{Y'}^2+2X'p_{X'}p_{Y'},
\qquad
\epsilon {\cal J}=p_{Y'}.
\end{gather*}
Contractions constructed from such limits would have the complex 2-sphere as the target manifold.
\item For this case it is most convenient to take the Hamiltonian in the form~\eqref{Darboux4}.
Let $x=x'$, $y=y'+\frac12\ln\epsilon$, ${\cal H}'= {\cal H}$, with~$b$ f\/ixed.
Then we have
\begin{gather*}
X_1'(x',y',p_{x'},p_{y'})=\frac{1}{\epsilon}X_1',
\qquad
X_2'(x',y',p_{x'},p_{y'})=\epsilon X_2'.
\end{gather*}
Contractions constructed from such limits would have $D4(b)$ as the target manifold.

\end{enumerate}

The manifold $D4(b)$ depends on a~parameter, so we can extend the contractions that we consider by allowing the
parameter to vary.
We f\/ind the following new contractions:
\begin{enumerate}\itemsep=0pt
\addtocounter{enumi}{3}
\item Let $X=\frac{X'}{2}-\frac12\ln(\epsilon)$, $Y=\frac{Y'}{2}$, $b=\frac{1}{\epsilon}$, ${\cal H}'=\frac{\epsilon}{2}
{\cal H}$.
Then as $\epsilon \to 0$ we have ${\cal H}'=\frac12 \frac{e^{2X'}}{e^{X'}+1}(p_{X'}^2+p_{Y'}^2)$, ${\cal
J}'=\frac{1}{2}{\cal J}=p_{Y'}$, and
\begin{gather*}
\frac{\epsilon}{2} {\cal Y}_1\approx -\cos(Y')\big({\cal H}'-e^{X'}p_{Y'}^2\big)-\sin(Y') e^{X'}p_{X'}^2 p_{Y'}^2,
\\
\frac{\epsilon}{2} {\cal Y}_2\approx -\sin(Y')\big({\cal H}'-e^{X'}p_{Y'}^2\big)+\cos(Y') e^{X'}p_{X'}^2 p_{Y'}^2.
\end{gather*}
Contractions constructed from such limits would have D3 as the target manifold.

\item Let $X=\frac{\epsilon X'}{2}$, $Y=\frac{\epsilon Y'}{2}$, $b=-2+\epsilon^2$, ${\cal H}'=\frac{\epsilon^2}{4} {\cal
H}$.
Then as $\epsilon \to 0$ we have ${\cal H}'\to \frac{{X'}^2}{{X'}^2+1}(p_{X'}^2+p_{Y'}^2)$, ${\cal
J}'=\frac{\epsilon}{2}{\cal J}=p_{Y'}$, and
\begin{gather*}
{\cal Y}_1 + {\cal H} - {\cal J}^2 - \frac{\epsilon^2}{2} {\cal H} \approx - 2 \left(2 X' Y' p_{X'} p_{Y'} +
\frac{\big({Y'}^2-{X'}^4\big)p_{Y'}^2 +{X'}^2 \big(1-{Y'}^2\big) p_{X'}^2}{{X'}^2+1} \right),
\\
{\cal Y}_2 \approx \frac{2}{\epsilon} \left(2X' p_{X'} p_{Y'} + \frac{2Y'}{{X'}^2+1}\big(p_{Y'}^2-{X'}^2 p_{X'}^2\big) \right).
\end{gather*}
Contractions constructed from such limits would have D2 as the target manifold.

\item Let $X=\epsilon X' +\frac{\ln(\epsilon)}{2}$, $Y=\epsilon Y'$, $b=-\frac{1}{\epsilon}$, ${\cal H}'=-2\epsilon^4{\cal H}$.
Then as $\epsilon \to 0$ we have ${\cal H}'\to \frac{1}{4X'}(p_{X'}^2+p_{Y'}^2)$, ${\cal J}'=\epsilon {\cal J}=p_{Y'}$, and
\begin{gather*}
{\cal Y}_1 + {\cal H} - \frac{1}{2\epsilon} {\cal J}^2 \approx \frac{1}{\epsilon^2} \left(p_{Y'}(Y'p_{X'} - X'p_{Y'}) -
\frac{{Y'}^2}{4X'}\big(p_{X'}^2+p_{Y'}^2\big) \right),
\\
{\cal Y}_2 \approx -\frac{1}{2\epsilon^3} \left(p_{X'} p_{Y'} - \frac{Y'}{2X'}\big(p_{X'}^2+p_{Y'}^2\big)\right).
\end{gather*}
Contractions constructed from such limits would have D1 as the target manifold.

\item Using Hamiltonian~\eqref{Darboux4b} we let $b=-2+\epsilon^2$, $X=X'$, $Y=Y'$, ${\cal H}\to {\cal H}'$.
Then
\begin{gather*}
{\cal H'}=\cosh^2 X'\big(p_{X'}^2+p_{Y'}^2\big),
\qquad
{\cal J}'={\cal J}=p_{Y'},
\\
{\cal Y}'_1=-\cos(2Y')\big({\cal H}'-\cosh 2X'p_{Y'}^2\big)-\sin (2Y')\sinh 2X'\, p_{X'}p_{Y'},
\\
{\cal Y}'_2=-\sin(2Y')\big({\cal H}'-\cosh 2X'p_{Y'}^2\big)+\cos (2Y')\sinh 2X'\, p_{X'}p_{Y'}.
\end{gather*}
Contractions constructed from such limits would have the complex 2-sphere as the target manifold.
Expressed in terms of the symmetries of the 2-sphere we have
\begin{gather*}
{\cal Y}'_1={\cal J}_1^2-{\cal J}_2^2,
\qquad
{\cal Y}'_2=2{\cal J}_1{\cal J}_2,
\qquad
{\cal J}'={\cal J}_3
\end{gather*}
\end{enumerate}

\subsection{Summary and examples of Darboux contractions}

From the results of~\cite{KM2014} and the preceding sections we have an analog of Theorem~\ref{theorem 4} for Darboux spaces:
\begin{theorem}
Every Lie algebra contraction of ${\cal G}3$ induces a~contraction of a~free geometric quadratic algebra $\tilde Q$ on
a~Darboux space, which in turn induces a~contraction of the quadratic algebra~$Q$ with potential.
This is true for both classical and quantum Darboux algebras.
\end{theorem}

Some examples follow.
\begin{enumerate}\itemsep=0pt
\item We describe how a~Lie algebra contraction induces the contraction of $D2C$ to $S4$, inclu\-ding the potential terms.
Recall for $D2C$ we have the symmetries~\eqref{D3Csym}, poten\-tial~\eqref{D3Cpot}, canonical equations~\eqref{D3Ccan} and
Casimir~\eqref{D3Ccas}.
The coordinate implementation is def\/ined by $x=\epsilon y'$, $y=\epsilon x'$.
In terms of the coordinates $x'$, $y'$ the canonical equations become
\begin{gather*}
{A'}^{12}=0,
\qquad
{A'}^{22}=-\frac{1}{y'},
\qquad
{B'}^{12}=\frac{2{x'}^2-{y'}^2}{y'\big({x'}^2+{y'}^2\big)},
\qquad
{b'}^{22}=-\frac{6x'}{\big({x'}^2+{y'}^2\big)}.
\end{gather*}
The contraction is def\/ined~by
\begin{gather*}
{\cal L'}_1={\cal L}_1,
\qquad
{\cal L'}_2=\frac{\epsilon}{2}{\cal L}_1,
\qquad
{\cal H}={\cal H}',
\qquad
{\cal R}'=\frac{\epsilon}{2}{\cal R}.
\end{gather*}
In the limit we f\/ind
\begin{gather*}
{\cal H}'={y'}^2\big(p_{x'}^2+p_{y'}^2\big)+V' ={\cal S}_2^2-2{\cal S}_1{\cal S}_3+V',
\\
{\cal L}'_1={\cal S}_2^2+W'_1,
\qquad
{\cal L}'_2={\cal S}_1{\cal S}_2+W'_2,
\end{gather*}
where
\begin{gather*}
V'=
\frac{{y'}^2}{2\sqrt{{x'}^2+{y'}^2}}\left(c_1+\frac{c_2}{\sqrt{{x'}^2+{y'}^2}+x'}+\frac{c_3}{\sqrt{{x'}^2+{y'}^2}-x'}\right)+c_4
\end{gather*}
and $b_1=\frac{c_1}{\epsilon}$, $b_2=c_2$, $b_3=c_3$, $b_4=c_4$.
Here
\begin{gather*}
{\cal S}_1=p_{x'},
\qquad
{\cal S}_2=x'p_{x'}+y'p_{y'},
\qquad
{\cal S}_3=\frac12\big({x'}^2-{y'}^2\big)p_{x'}^2+x'y'p_{y'},
\end{gather*}
are the 1st order generators of the symmetry algebra for the free Hamiltonian ${\cal H}'_0 ={y'}^2(p_{x'}^2+p_{y'}^2)$
on the Poincar\'e upper half plane: $x$  real, $y>0$.
The Casimir becomes in the limit:
\begin{gather*}
{{\cal R}'}^2-4{\cal L}_1{'{\cal L}'_2}^2+(c_2+c_3){\cal L}'_2-\frac{c_1^2}{4}{\cal H}' +\frac{c_1^2}{4}{\cal
L}'_1\\
\qquad{} +\frac{c_1}{2}(c_2-c_3){\cal L}'_2+\frac{{c'_1}^2}{16}(c_2+c_3+4c_4)=0.
\end{gather*}

\item${D4(b)A'\to D3E}$: We again give more details in our 2nd example, which involves changing the parameter~$b$.
For the degenerate system
\begin{gather*}
{\cal H} =\frac{\sinh^2(2x)}{2\cosh(2x)+b}\big(p_x^2+p_y^2\big)+\frac{\alpha}{2\cosh(2x)+b},
\end{gather*}
we can get
\begin{gather*}
{\cal H'}=\frac12\frac{e^{2x'}}{e^{x'}+1}\big(p_{x'}^2+p_{y'}^2\big)+\frac{\beta}{e^{x'}+1}
\end{gather*}
as a~contraction case by taking
\begin{gather*}
x=\frac{x'}{2}-\frac12\ln(\epsilon),
\qquad
y=\frac{y'}{2},
\qquad
b=\frac{1}{\epsilon},
\qquad
{\cal H}'=8\epsilon{\cal H},
\qquad
\alpha=\frac{\beta}{8\epsilon^2}.
\end{gather*}
Then we have ${\cal J}'=2{\cal J}=p_{y'}$, and ${\cal Y}'_1\approx -4\epsilon {\cal Y}_1$, ${\cal Y}'_2\approx
-4\epsilon {\cal Y}_2$.

\item ${\tilde D1A \to \tilde E9(a=-1)}$: For system ${\tilde D1A}$ we have
\begin{gather*}
{\cal L}_1={\cal X}_2+b{\cal K}^2,
\qquad
{\cal L}_2={\cal X}_1+b{\cal K}^2,
\\
{\cal R}^2=2i{\cal L}_2^3+8i{\cal L}_1{\cal L}_2{\cal H}+4b{\cal L}_2^2{\cal H}+16b{\cal H}^2{\cal L}_1.
\end{gather*}
We choose contraction 1) for $\tilde D1A$: $x=x'+\frac{1}{\epsilon}$, $y=y'$, ${\cal H}=\frac{\epsilon}{4}{\cal H}'$ and
let $b=\frac{1}{\epsilon}$.
Then as $\epsilon\to 0$ we f\/ind
\begin{gather*}
{\cal L}_1\to {\cal L}_1'=-p_{y'}(y'p_{x'}-x'p_{y'}),
\qquad
{\cal L}_2\to {\cal L}_2'=p_{y'}(p_{x'}+ip_{y'}),
\qquad
{\cal H}'= p_{x'}^2+p_{y'}^2,
\end{gather*}
and with ${\cal R}'=\{{\cal L}_1',{\cal L}_2'\}$ the Casimir becomes ${{\cal R}'}^2=2i{{\cal L}'_2}^3+{{\cal
L}_2'}^2{\cal H}'$.

\item ${\tilde D1B \to \tilde E2}$: For system ${\tilde D1B}$ we have
\begin{gather*}
{\cal L}_1={\cal X}_2,
\qquad
{\cal L}_2={\cal K}^2,
\qquad
{\cal R}^2=-4{\cal L}_2^3-16{\cal L}_1{\cal L}_2{\cal H}.
\end{gather*}
We choose contraction 1) for ${\tilde D1B}$: $x=x'+\frac{1}{\epsilon}$, $y=y'$, ${\cal H}=\frac{\epsilon}{4}{\cal H}'$.
Then as $\epsilon\to 0$ we f\/ind
\begin{gather*}
{\cal L}_1\approx {\cal L}_1'-\frac{1}{\epsilon}{\cal L}'_2,
\qquad\!
{\cal L}'_1=-p_{y'}(x'p_{y'}-y'p_{x'}),
\qquad\!
{\cal L}_2= {\cal L}_2'=p_{y'}^2,
\qquad\!
{\cal H}'= p_{x'}^2+p_{y'}^2,
\end{gather*}
and with ${\cal R}'=\{{\cal L}_1',{\cal L}_2'\}$ the Casimir becomes ${{\cal R}'}^2=-4{{\cal L}'_2}^3+4{{\cal
L}_2'}^2{\cal H}'$.

\item ${\tilde D1C \to \tilde E3'}$: For system ${\tilde D1C}$ we have ${\cal L}_1={\cal X}_1$, ${\cal L}_2={\cal K}^2$,
$ {\cal R}^2=16{\cal L}_2{\cal H}^2$.
We choose contraction 1) for $\tilde D1C$: $x=x'+\frac{1}{\epsilon}$, $y=y'$, ${\cal H}=\frac{\epsilon}{4}{\cal H}'$.
Then as $\epsilon\to 0$ we f\/ind
\begin{gather*}
{\cal L}_1\to {\cal L}_1'=p_{x'}'p_{y'},
\qquad
{\cal L}_2= {\cal L}_2'=p_{y'}^2,
\qquad
{\cal H}'= p_{x'}^2+p_{y'}^2,
\end{gather*}
and with ${\cal R}'=\{{\cal L}_1',{\cal L}_2'\}$ the Casimir becomes ${{\cal R}'}^2=0$.
\item $\tilde D4(b)C'\to \tilde S7$ as $b\to -2$: For system $\tilde D4(b)C'$ we have
\begin{gather*}
{\cal L}1={\cal Y}_1-{\cal J}^2,
\qquad
{\cal L}_1={\cal Y}_2,
\\
{\cal R}^2=4b{\cal H}^3-8{\cal L}_13-8{\cal L}_1{\cal L}_2^2+4b\big({\cal L}_1^2+{\cal L}_2^2\big){\cal H}+8{\cal L}_1{\cal H}^2.
\end{gather*}
In the limit as $b\to -2$ we have
\begin{gather*}
{\cal L}_1'=2{\cal J}_1^2-{\cal H}',
\qquad
{\cal L}_2'=2{\cal J}_1{\cal J}_2,
\end{gather*}
where ${\cal J}_1$, ${\cal J}_2$, ${\cal J}_3$ are the symmetries of the sphere and ${\cal H}'={\cal J}_1^2+{\cal J}_2^2+{\cal J}_3^2$.
The limit system is $\tilde S7$.
\end{enumerate}

\section{Tables/f\/igures describing the contractions\\ of the nondegenerate and degenerate Darboux systems}

We now tabulate the contractions of Darboux systems.
The system on the left is the starting system, and an asterisk indicates that a~system is degenerate

\subsection{D1 contraction table}

{\renewcommand{\arraystretch}{1.5}
\begin{center}
\begin{tabular}
{c | c c c} & 1: $D1$ to f\/lat space & 2: $D1$ to $D1$ & 3: $D1$ to $D1$
\\
\hline
$\tilde D1A$ & $\tilde E3'$, $\tilde E9$ & $\tilde D1C$, $\tilde D1A$ & $\tilde D1C$, $\tilde D1A$
\\
$\tilde D1B$ & $\tilde E2$ & $\tilde D1C$ & $\tilde D1B$
\\
$\tilde D1C$ & $\tilde E3'$ & $\tilde D1C$ & $\tilde D1C$
\\
$\tilde D1D^*$ & $\tilde E5^*$ & $\tilde D1D^*$ & $\tilde D1D^*$
\\
\end{tabular}
\end{center}

}

\subsection{D2 contraction table}

{\renewcommand{\arraystretch}{1.5}
\begin{center}
\begin{tabular}{c | c c c} & 1: $D2$ to f\/lat space & 2: $D2$ to 2-sphere & 3: $D2$ to $D2$
\\
\hline
$\tilde D2A$ & $\tilde E3'$ & $\tilde S1$ & $\tilde D2A$
\\
$\tilde D2B$ & $\tilde E2$ & $\tilde S2$ & $\tilde D2A$
\\
$\tilde D2C$ & $\tilde E3'$ & $\tilde S4$ & $\tilde D2A$
\\
$\tilde D2D^*$ & $\tilde E5^*$ & $\tilde S5^*$ & $\tilde D2D^*$
\\
\end{tabular}
\end{center}}

\subsection{D3 contraction table}

{\renewcommand{\arraystretch}{1.5}
\begin{center}
\begin{tabular}
{c | c c c} & 1: $D3$ to f\/lat space & 2: $D3$ to f\/lat space & 3: $D3$ to $D3$
\\
\hline
$\tilde D3A$ & $\tilde E3'$ & $\tilde E3'$ & $\tilde D3D$
\\
$\tilde D3B$ & $\tilde E3'$ & $\tilde E3'$ & $\tilde D3B$
\\
$\tilde D3C$ & $\tilde E1$ & $\tilde E2$ & $\tilde D3D$
\\
$\tilde D3D$ & $\tilde E8$ & $\tilde E3'$ & $\tilde D3D$
\\
$\tilde D3E^*$ & $\tilde E3^*$ & $\tilde E5^*$ & $\tilde D3E^*$
\\
\end{tabular}
\end{center}}

\subsection{D4 contraction table}

{\renewcommand{\arraystretch}{1.5}
\begin{center}
\begin{tabular}
{c | c c c} & 1: $D4$ to f\/lat space & 2: $D4$ to 2-sphere & 3: $D4$ to $D4$
\\
\hline
$\tilde D4(b)A$ & $\tilde E1$ & $\tilde S2$ & $D4(b)B$
\\
$\tilde D4(b)B$ & $\tilde E8$ & $\tilde S1$ & $D4(b)B$
\\
$\tilde D4(b)C$ & $\tilde E3'$ & $\tilde S4$ & $D4(b)B$
\\
$\tilde D4(b)D^*$ & $\tilde E3^*$ & $\tilde S5^*$ & $D4(b)D^*$
\\
\end{tabular}

\vspace{3mm}

\begin{tabular}
{c | c c c c} & 4: $D4$ to $D3$ & 5: $D4$ to $D2$ & 6: $D4$ to $D1$ & 7: $D4$ to 2-sphere
\\
\hline
$\tilde D4(b)A$ & $\tilde D3C$ & $\tilde D2B$ & $\tilde D1B$ &$\tilde S9$
\\
$\tilde D4(b)B$ & $\tilde D3D$ & $\tilde D2A$ & $\tilde D1C$ &$\tilde S2$
\\
$\tilde D4(b)C$ & $\tilde D3B$ & $\tilde D2C$ & $\tilde D1C$ & $\tilde S7$
\\
$\tilde D4(b)D^*$ & $\tilde D3E^*$ & $\tilde D2D^*$ & $\tilde D1D^*$ & $\tilde S3^*$
\\
\end{tabular}
\end{center}
}

\subsection{Free nondegenerate Darboux systems to degenerate systems}

The following are not contractions in the usual sense, but we show which nondegenerate systems restrict to degenerate ones.

\begin{center}
\begin{tabular}
{c | c} nondegenerate &$\to$ degenerate
\\
\hline
${\tilde D}1B $& ${\tilde D}1D^*$
\\
${\tilde D}1C $& ${\tilde D}1D^*$
\\
${\tilde D}2A $& ${\tilde D}2D^*$
\\
${\tilde D}2B$ &${\tilde D}2D^*$
\\
${\tilde D}3C$ & ${\tilde D}3E^*$
\\
${\tilde D}3D$ & ${\tilde D}3E^*$
\\
${\tilde D}4(b)A$ & ${\tilde D}4(b)D^*$
\\
${\tilde D}4(b)B$ & ${\tilde D}4(b)D^*$
\\
\end{tabular}
\end{center}

\subsection{Figures}

What follows are diagrams illustrating the contractions of Darboux systems.
Boxes represent systems (blue boxes being nondegenerate and red boxes being degenerate), and arrows represent
contractions.
In these diagrams, certain contractions that do not af\/fect the overall hierarchy have been omitted for aesthetic
reasons, and a~dotted arrow indicates that the limiting process inducing the contraction changes the free Hamiltonian.
A~diagram illustrating the nondegenerate Darboux systems which restrict to degenerate ones is also presented.

\begin{figure}
[t] \centering \includegraphics[scale=0.51]{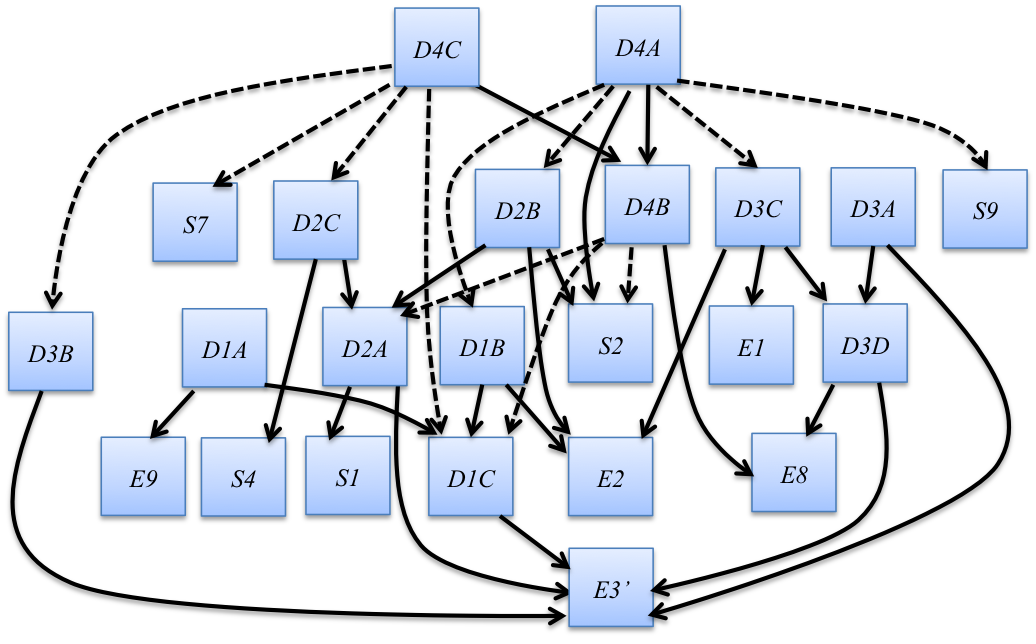} \caption{Diagram indicating the contractions of
nondegenerate Darboux systems.}
\end{figure}

\begin{figure}
[t] \centering \includegraphics[scale=0.51]{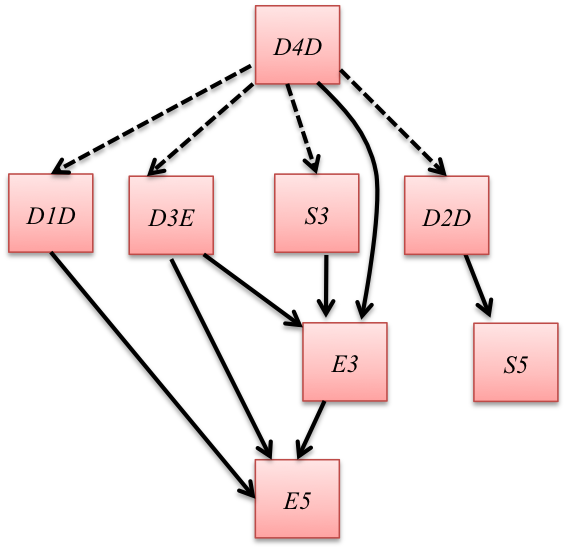} \caption{Diagram indicating the contractions of
degenerate Darboux systems.}
\end{figure}

\begin{figure}
[t] \centering \includegraphics[scale=0.51]{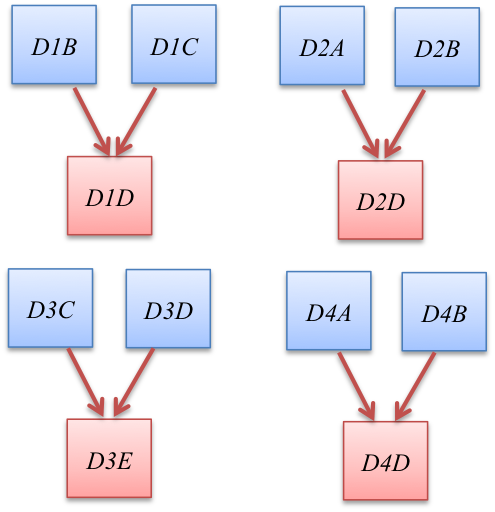} \caption{Diagram indicating restrictions of nondegenerate
Darboux systems to degenerate ones.}
\end{figure}

For completeness we give the def\/initions of the constant curvature superintegrable systems that are targets of Darboux
contractions.
The following lists contain the def\/ining relations for the free systems and the potentials of the superintegrable
systems.

{\bf Degenerate Euclidean targets}: Here the coordinates are $x$, $y$ and $p_1=p_x$, $p_2=p_y$, and ${\cal
J}=xp_2-yp_1$.
\begin{enumerate}\itemsep=0pt
\item $\tilde E3$: ${\cal H}=p_1^2+p_2^2$, ${\cal X}= {\cal J}$, ${\cal L}_1=p_1^2$, ${\cal L}_2=p_1p_2$, \\
Casimir: $-{\cal L}_2^2-{\cal L}_1({\cal L}_1-{\cal H})=0$, potential: $V=\alpha(x^2+y^2)$.  

\item $\tilde E5$:\ ${\cal H}=p_1^2+p_2^2$, ${\cal X}= p_1$, ${\cal L}_1={\cal J}{p_1}$, ${\cal
L}_2=p_2{p_1}$,\\ Casimir: $\frac12({\cal L}_2^2+{\cal X}^4-{\cal H}{\cal X}^2)=0$, potential: $
V=\alpha x$.
\end{enumerate}

{\bf Nondegenerate Euclidean targets}:
\begin{enumerate}\itemsep=0pt
\item $\tilde E1$:
\qquad
${\cal L}_1={\cal J}^2$, ${\cal L}_2=p_1^2$, ${\cal R}^2=16{\cal L}_1{\cal L}_2({\cal H}-{\cal L}_2)$,
$V=\alpha(x^2+y^2)+\frac{\beta}{x^2}+\frac{\gamma}{y^2}$,

\item $\tilde E2$:
\qquad
${\cal L}_1=p_2^2$, ${\cal L}_2=p_2{\cal J}$, ${\cal R}^2=4{\cal L}_1^2({\cal H}-{\cal L}_1)$, $V=\alpha(4x^2+y^2)+\beta
x+\frac{\gamma}{y^2}$,

\item $\tilde E3'$:
\qquad
${\cal L}_1=p_1^2$, ${\cal L}_2=p_1p_2$, ${\cal R}^2=0$, $V=\alpha(x^2+y^2)+\beta x+\gamma y$,

\item $\tilde E8$:
\qquad
${\cal L}_1={\cal J}^2$, ${\cal L}_2=(p_1+ip_2)^2$, ${\cal R}^2=-16{\cal L}_1{\cal L}_2^2$, $V=\frac{\alpha
(x-iy)}{(x+iy)^3}+\frac{\beta}{(x+iy)^2}+\gamma(x^2+y^2)$,

\item $\tilde E9$:
\qquad
${\cal L}_1=(p_1+ip_2)^2$, ${\cal L}_2=p_1{\cal J}$, ${\cal R}^2=-2{\cal L}_1({2\cal L}_1+{\cal H})^2$,
$V=\frac{\alpha}{\sqrt{x+iy}}+\beta y+\frac{\gamma (x+2iy)}{\sqrt{x+iy}}$,
\end{enumerate}

{\bf Degenerate targets on the 2-sphere}: Here ${\cal J}_1=yp_3-zp_2$, ${\cal J}_2=zp_1-xp_3$, ${\cal
J}_3=xp_2-yp_1$.
\begin{enumerate}\itemsep=0pt
\item $\tilde S3$:
${\cal H}={\cal J}_1^2+{\cal J}_2^2+{\cal J}_3^2$, ${\cal X}= {\cal J}_3$, ${\cal L}_1=({\cal
J}_1+i{\cal J}_2)^2$, ${\cal L}_2=({\cal J}_1-i{\cal J}_2)^2$,\\ Casimir: $-2i(({\cal H}-{\cal
X}^2)^2-{\cal L}_1{\cal L}_2)=0$, potential: $V=\frac{\alpha}{z^2}$,
\item $\tilde S5$:
${\cal H}={\cal J}_1^2+{\cal J}_2^2+{\cal J}_3^2$, ${\cal X}= {\cal J}_1+i{\cal J}_2$, ${\cal
L}_1={\cal J}_3^2$, ${\cal L}_2=({\cal J}_1+i{\cal J}_2){\cal J}_3$,\\ Casimir: $-i({\cal L}_2^2-{\cal
X}^2{\cal L}_1)=0$, potential: $V=\frac{\alpha}{(x+iy)^2}$.
\item $\tilde S6$:
${\cal H}={\cal J}_1^2+{\cal J}_2^2+{\cal J}_3^2$, ${\cal X}= {\cal J}_3$, ${\cal L}_1={\cal
J}_3{\cal J}_1$, ${\cal L}_2={\cal J}_3{\cal J}_2$,\\  Casimir: $-\frac12({\cal L}_1^2+{\cal
L}_2^2+{\cal X}^2({\cal X}^2-{\cal H}))=0$, potential: $V=\frac{\alpha z}{\sqrt{x^2+y^2}}$,
\end{enumerate}

{\bf Nondegenerate targets on the 2-sphere}:
\begin{enumerate}\itemsep=0pt
\item $\tilde S1$:\quad
${\cal L}_1= ({\cal J}_1+i{\cal J}_2){\cal J}_3$, ${\cal L}_2=({\cal J}_1+i{\cal J}_2)^2$, ${\cal
R}^2=-4{\cal L}_2^3$, $V=\frac{\alpha}{(x+iy)^2}+\frac{\beta z}{(x+iy)^2}+\frac{\gamma(1-4z^2)}{(x+iy)^4}$,
\item $\tilde S2$:\quad
${\cal L}_1= ({\cal J}_1+i{\cal J}_2)^2$, ${\cal L}_2={\cal J}_3^2$, ${\cal R}^2=-16{\cal L}_1^2{\cal L}_2$,
$V=\frac{\alpha}{z^2}+\frac{\beta}{(x+iy)^2}+\frac{\gamma(x-iy)}{(x+iy)^3}$,
\item $\tilde S4$:\quad
${\cal L}_1={\cal J}_3^2$, ${\cal L}_2=({\cal J}_1+i{\cal J}_2){\cal J}_3$, ${\cal R}^2=-4{\cal L}_1{\cal L}_2^2$,
$V=\frac{\alpha}{(x+iy)^2}+\frac{\beta z}{\sqrt{x^2+y^2}}+\frac{\gamma}{(x+iy)\sqrt{x^2+y^2}}$,
\item $\tilde S7$:\quad
${\cal L}_1={\cal J}_3^2$, ${\cal L}_2={\cal J}_1{\cal J}_3$, ${\cal R}^2=-4{\cal L}_1^3-4{\cal
L}_2^2{\cal L}_1+4{\cal L}_1^2{\cal H}$, $V=\frac{\alpha z}{\sqrt{x^2+y^2}}+\frac{\beta
x}{y^2\sqrt{x^2+y^2}}+\frac{\gamma}{y^2}$,
\item $\tilde S9$:\quad
${\cal L}_1={\cal J}_3^2$, ${\cal L}_2={\cal J}_1^2$, ${\cal R}^2=-16{\cal L}_1^2{\cal L}_2-16{\cal L}_1{\cal
L}_2^2+16{\cal L}_1{\cal L}_2{\cal H}$, $V=\frac{\alpha}{x^2}+\frac{\beta}{y^2}+\frac{\gamma}{z^2}$,
\end{enumerate}

\section{Conclusions and discussion}

This paper is part of a~series on 2D quadratic algebras, their classif\/ication, structure, representations, and
especially, contractions as they relate to 2nd order 2D superintegrable systems.
Of special interest are contractions that correspond to geometrical pointwise limiting processes in the physical
superintegrable systems.
As shown in~\cite{KMP2014}, one of the consequences of contrac\-ting between superintegrable systems is a~series of
limiting relations between special functions associated with the superintegrable systems, a~special case of which is the
Askey scheme for hypergeometric orthogonal polynomials.
In~\cite{KM2014} we studied quadratic algebras related to 2nd order superintegrable systems on constant curvature spaces
and showed that there is a~one-to-one correspondence between conjugacy classes of quadratic algebras in the enveloping
algebras of $e(2,{\mathbb C})$ and $o(3,{\mathbb C})$, and isomorphism classes of superintegrable systems with
potential.
Further, we showed for constant curvature spaces that generalizations of In\"on\"u--Wigner Lie algebra contractions of
$e(2,{\mathbb C})$ and $o(3,{\mathbb C})$, induce quadratic algebra contractions that correspond to geometrical
pointwise limiting processes in the physical systems.
The procedure is rigid and deterministic.
The present paper extends these results and shows that Darboux superintegrable systems are also characterized by free
quadratic algebras contained in the symmetry algebras of these spaces and that their contractions are also induced~by
In\"on\"u--Wigner contractions.
Thus our basic results hold for all 2nd order 2D superintegrable systems.
In follow-up papers, in preparation, we will classify abstract quadratic algebras and their contractions, including
those not induced from Lie algebras, and study which of these relate to superintegrable systems.

We intend to conclude this series by relating contractions of 2nd order superintegrable systems to limiting processes
for $R$-separable coordinate systems for wave equations, introduced by B\^ocher in his famous 1894 thesis~\cite{Bocher}.
We will show that in 2D, B\^ocher's limiting processes for cyclides and ellipses induce generalizations of
In\"on\"u--Wigner contractions of the $so(4,{\mathbb C})$ conformal symmetry algebra of the 2D wave equation with
potential and that these contractions explain the full contraction pattern for 2nd order superintegrable systems.
B\^ocher's limits, which we term B\^ocher contractions, apply to all dimensions $n\ge 2$, so this should provide
a~useful guide to the analysis of 2nd order superintegrable systems in higher dimensions.

\subsection*{Acknowledgment}
This work was partially supported by a~grant from the Simons Foundation (\# 208754 to Willard Miller, Jr.).

\pdfbookmark[1]{References}{ref}
\LastPageEnding

\end{document}